\documentclass[aps,prb,reprint,twocolumn,superscriptaddress,floatfix,longbibliography,nofootinbib]{revtex4-1}
\usepackage{epsfig,amsmath,amssymb,color,comment,physics}
\usepackage[makeroom]{cancel}
\usepackage[caption=false]{subfig}
\usepackage{mathrsfs}
\usepackage[countmax]{subfloat}
\usepackage[normalem]{ulem}
\usepackage[english]{babel}
\usepackage{dsfont}
\usepackage{placeins}
\usepackage{multirow}
\usepackage[bookmarks=true,colorlinks,linkcolor=NavyBlue,urlcolor=NavyBlue,citecolor=RoyalBlue]{hyperref}
\usepackage[dvipsnames]{xcolor}

\usepackage{color}

\usepackage[pagewise]{lineno}

\begin{document}

\title{Driving quantum many-body scars in the PXP model}

\author{Ana Hudomal}
 \affiliation{School of Physics and Astronomy, University of Leeds, Leeds LS2 9JT, UK}
  \affiliation{Institute of Physics Belgrade, University of Belgrade, 11080 Belgrade, Serbia}

\author{Jean-Yves Desaules}
 \affiliation{School of Physics and Astronomy, University of Leeds, Leeds LS2 9JT, UK}
 
 \author{Bhaskar Mukherjee}
\affiliation{Department of Physics, University College London, Gower Street, London WC1E 6BT, UK}
 
\author{Guo-Xian Su}
 \affiliation{Hefei National Laboratory for Physical Sciences at Microscale and Department of Modern Physics, University of Science and Technology of China, Hefei, Anhui 230026, China}
 \affiliation{Physikalisches Institut, Ruprecht-Karls-Universit\"{a}t Heidelberg, Im Neuenheimer Feld 226, 69120 Heidelberg, Germany}
 \affiliation{
	CAS Center for Excellence and Synergetic Innovation Center in Quantum Information and Quantum Physics,
	University of Science and Technology of China, Hefei, Anhui 230026, China
}

\author{Jad C.~Halimeh}
\affiliation{Department of Physics and Arnold Sommerfeld Center for Theoretical Physics (ASC), Ludwig-Maximilians-Universit\"at M\"unchen, Theresienstra\ss e 37, D-80333 M\"unchen, Germany}
\affiliation{Munich Center for Quantum Science and Technology (MCQST), Schellingstra\ss e 4, D-80799 M\"unchen, Germany}

\author{Zlatko Papi\'c}
\affiliation{School of Physics and Astronomy, University of Leeds, Leeds LS2 9JT, UK}

\begin{abstract}
Periodic driving has been established as a powerful technique for engineering novel phases of matter and intrinsically out-of-equilibrium phenomena such as time crystals. Recent work by Bluvstein \emph{et al.}~[\href{https://www.science.org/doi/10.1126/science.abg2530}{Science {\bf 371}, 1355 (2021)}] has demonstrated that periodic driving can also lead to a significant enhancement of quantum many-body scarring, whereby certain non-integrable systems can display persistent quantum revivals from special initial states. Nevertheless, the mechanisms behind driving-induced scar enhancement remain poorly understood. Here we report a detailed study of the effect of periodic driving on the PXP model describing Rydberg atoms in the presence of a strong Rydberg blockade -- the canonical static model of quantum many-body scarring. We show that periodic modulation of the chemical potential gives rise to a rich phase diagram, with at least two distinct types of scarring regimes that we distinguish by examining their Floquet spectra. We formulate a toy model, based on a sequence of square pulses, that accurately captures the details of the scarred dynamics and allows for analytical treatment in the large-amplitude and high-frequency driving regimes. Finally, we point out that driving with a spatially inhomogeneous chemical potential allows to stabilize quantum revivals from arbitrary initial states in the PXP model, via a mechanism similar to prethermalization. 

\end{abstract}

\date{\today}
\maketitle
\tableofcontents

\section{Introduction}\label{s:intro}
Quantum many-body scarring (QMBS) \cite{Serbyn2021} has recently been added to the growing  list of ergodicity-breaking phenomena, previously actively studied in integrable models~\cite{sutherland2004beautiful} and many-body localization~\cite{Huse-rev, AbaninRMP}. In conventional ergodic systems, even in the absence of a coupling to a thermal bath, the interactions between constituent degrees of freedom lead to fast thermalization~\cite{DeutschETH,SrednickiETH,RigolNature}. A distinguishing feature of QMBS systems is that ergodicity is broken by a relatively small subset of the system's energy eigenstates, which coexist with the thermalizing bulk of the spectrum~\cite{Serbyn2021,MoudgalyaReview}. Experimentally, QMBS can be detected by preparing the system in special initial states that have predominant support on the ergodicity-breaking eigenstates. Such initial states are found to  display persistent quantum revivals, in contrast to other ``generic" initial states whose dynamics is featureless~\cite{Bernien2017,Bluvstein2021}. The investigations of such ``weak" ergodicity breaking phenomena continue to attract much attention, both in terms of new experimental realizations~\cite{Kao2021,Scherg2020,Jepsen2021} as well as universal 
mechanisms~\cite{BernevigEnt, ShiraishiMori, MotrunichTowers, Dea2020, Pakrouski2020,Ren2021} for violating the Eigenstate Thermalization Hypothesis (ETH)~\cite{DeutschETH,SrednickiETH}. Theoretical studies continue to identify large classes of models that exhibit various aspects of QMBS phenomena. Some notable examples include various nonintegrable lattice models~\cite{Iadecola2019_2,Iadecola2019_3,Bull2019,Chattopadhyay,OnsagerScars,MoudgalyaFendley,SuraceSUSY,Kuno2020}, models of correlated fermions and bosons~\cite{Vafek,MarkHubbard,MoudgalyaHubbard,Moudgalya2019,bosonScars,Zhao2020,Desaules2021,Pakrouski2021,Su2022}, frustrated magnets~\cite{Lee2020,McClarty2020}, topological phases of matter \cite{NeupertScars,Wildeboer2020}, lattice gauge theories \cite{Surace2019,Banerjee2021,biswas2022scars,Desaules2022,Desaules2022prominent}, and periodically-driven systems \cite{Buca2019,BucaHubbard,Sugiura2019,Mizuta2020,Mukherjee2020a,Mukherjee2020b,Haldar2021}.

Recent experiments\cite{Bluvstein2021} on QMBS utilising Rydberg atom arrays in various geometries, both one-dimensional and two-dimensional, have shown that periodic driving can stabilize the revivals that occur in the static system~\cite{Turner2018a,Turner2018b,wenwei18TDVPscar}. Similar effect was also observed in experiments on a Bose-Hubbard quantum simulator~\cite{Su2022}, where QMBS phenomenon was identified in a larger class of  initial states. In both cases, the optimal driving frequency was empirically found to be related to the frequency of revivals in the static model. Although the driving protocol is relatively simple,  the mechanism behind it remains to be understood. Previous work in Ref.~\onlinecite{Maskara2021} provided an important insight that the experimental driving protocol can be approximated by a simpler two-step (kicked) driving scheme. Such a toy model is easier to treat analytically and it captures some of the observed phenomenology. However, its relation to the experimental driving protocol is not transparent, in particular there is no direct mapping between the parameters of the two protocols (except for the driving frequency, which is equal in the two cases).

Moreover, further questions naturally arise. For example, the only two initial states whose revivals were stabilized by periodic driving (the N\'eel and polarized states) are precisely the ones that exhibit QMBS in the static version of the same model (the polarized state becomes scarred in the presence of static detuning~\cite{Su2022}). This suggests that there are fundamental limitations to the driving, which only ``works" for such special states. If the enhancement mechanism is indeed related to QMBS, it would imply that it should be possible to improve the revivals in a variety of other QMBS models that have been studied theoretically. Another question is how to predict  the optimal driving parameters without performing a computationally-costly brute force search. To this end, Ref.~\onlinecite{Bluvstein2021} has identified a window of driving frequencies centered around twice the frequency of revivals in the static model. A subharmonic response to driving was observed in this frequency regime, suggesting a relation to the discrete time crystal, which also warrants further investigation. 

In this work we numerically and analytically study the effects of driving on the PXP model describing Rydberg atoms in the regime of strong Rydberg blockade. In Sec.~\ref{s:cosine} we introduce the model and investigate in detail the spatially-uniform cosine driving scheme that was used in recent experiments. We identify two regimes with distinct properties (dubbed Type-1 and Type-2). In Sec.~\ref{s:revivals} we focus on the optimally driven N\'eel and polarized state. In Sec.~\ref{s:square} we introduce a toy model based on the square pulse driving protocol. We show that this protocol better approximates the cosine-drive dynamics than the previously studied kicked drive~\cite{Maskara2021}. The square-pulse protocol is amenable to analytical treatment in certain limits, as we demonstrate for the high-amplitude and high-frequency drive regimes. Finally, in Sec.~\ref{s:scars} we discuss the implications of our results on the relation between QMBS and the enhancement by periodic driving, suggesting that QMBS is a necessary ingredient in the Type-1 (typically low-amplitude) regime, while Type-2 driving (typically high-amplitude) can be also applied to non-scarred models. We summarize our findings and identify some future directions in Sec.~\ref{s:conclusions}.
Appendices~\ref{a:scaling}-\ref{a:trajectory_polarized} provide additional information on system size scaling, a modified (spatially inhomogeneous) cosine driving protocol that can generate revivals from \emph{any} initial product state, high frequency expansion for the Floquet Hamiltonian, and an alternative way to visualize the trajectory of a state in Hilbert space.

\section{PXP model and cosine drive}\label{s:cosine}

The PXP model~\cite{FendleySachdev,Lesanovsky2012} describes a kinetically constrained chain of spin-$1/2$ degrees of freedom. The chain serves as a model for the system of Rydberg atoms, each of which can exist in two possible states,  $\ket{\circ}$ and  $\ket{\bullet}$, corresponding to the ground state and the excited state, respectively. An array of $N$ such atoms is governed by the Hamiltonian 
\begin{align}\label{Eq:PXP}
	H_{\mathrm{PXP}} = \Omega \sum_{j=1}^N P_{j-1} X_j P_{j+1},
\end{align}
where $X_j = {\ket{\circ_j}} {\bra{\bullet_j}} +  {\ket{\bullet_j}} {\bra{\circ_j}}$ is the Pauli $x$ matrix on site $j$, describing local Rabi precession with frequency $\Omega$. The projectors onto the ground state, $P_j  = \ket{\circ_j}\bra{\circ_j}$, constrain the dynamics by allowing an atom to flip its state only if both of its neighbors are in the ground state. Unless specified otherwise, we use periodic boundary conditions (PBC) and set the PXP term strength to $\Omega=1$.

Before introducing the driving scheme, we briefly review some pertinent properties of the static PXP model. The PXP model is quantum chaotic~\cite{Turner2018a}. However, it was found that preparing the system in the N\'eel or $\ket{\mathbb{Z}_2} \equiv \ket{ \bullet\circ\bullet\circ...\bullet\circ}$ state, which has an excitation on every other site, leads to  persistent quantum revivals~\cite{Sun2008, Olmos2010,Turner2018b}. Analogously, the translated N\'eel state, $\ket{\mathbb{Z}_2'} \equiv \ket{ \circ\bullet\circ\bullet...\circ\bullet}$, also displays revivals. In a conventional thermalizing system, the revivals are not expected as the $\ket{\mathbb{Z}_2}$ initial state has predominant support on the energy eigenstates in the middle of the many-body spectrum, thus it corresponds to ``infinite" temperature.
The presence of revivals in a special initial state in an overall chaotic system was understood to be a many-body analog of the phenomena associated with a single particle inside a stadium billiard, where nonergodicity arises as a ``scar" imprinted by a particle's classical periodic orbit~\cite{Heller84, wenwei18TDVPscar,Turner2021}. 
In QMBS systems, the eigenstates were shown to form tower structures~\cite{Turner2018a}, resulting from the anomalously high overlap of eigenstates with the $\ket{\mathbb{Z}_2}$ initial state, and the equal energy spacing between the towers leads to  quantum revivals.
We mention that the scarring in the PXP model was also shown to persist in higher dimensions~\cite{Michailidis2D,Hsieh2020,Bluvstein2021} and in the presence of certain perturbations \cite{Turner2018b,Khemani2018,Lin2020}, including disorder~\cite{MondragonShem2020}.

The revivals in the PXP model are not perfect, in particular the wave function fidelity, 
\begin{eqnarray}
F(t) \equiv |\langle \psi (t) |  \psi_0\rangle |^2,
\end{eqnarray}
with the initial state chosen to be $\ket{\psi_0} = \ket{\mathbb{Z}_2}$, visibly decays over moderate times~\cite{Turner2018b}.  Reference~\onlinecite{Bluvstein2021} showed that the decay time-scale can be significantly extended by introducing a cosine modulation $\Delta(t)$ of the chemical potential, 
\begin{eqnarray}
\label{eq:ht}
H(t) = H_\mathrm{PXP}  -  \Delta(t) \sum_i n_{i}, \;\;
\Delta(t) = \Delta_0+\Delta_m\cos(\omega t), \;\;\;\;\;\ \label{eq:cos_drive}
\end{eqnarray}
where $n_i = \ket{\bullet_i}\bra{\bullet_i}$ is the density of excitations on site $i$, $\Delta_0$ is the static detuning, $\Delta_m$ modulation amplitude and $\omega$ the driving frequency.  

The periodic modulation of the chemical potential in Eq.~(\ref{eq:cos_drive}) was shown to stabilize and enhance the revivals in the quantum fidelity, both in numerical simulations and in experiments on quantum simulators~\cite{Su2022}. The enhancement was achieved for two different initial states, the previously mentioned  $\ket{\mathbb{Z}_2}$ state, as well as the polarized state, which has no excitations,  $\ket{0} \equiv \ket{{\circ}{\circ}{\circ}\ldots{\circ}{\circ}{\circ}}$. 
The optimal driving frequency for the N\'eel state was found to be close to twice the frequency of revivals in the pure PXP model [$\Delta(t)=0$]; for the polarized state, the optimal frequency was instead close to the frequency of revivals in the detuned PXP model [$\Delta(t)=\Delta_0$]. The optimal values of $\Delta_0$ and $\Delta_m$ were found empirically and have different values for these two states. 

\begin{figure}[htb]
 \includegraphics[width=0.43\textwidth]{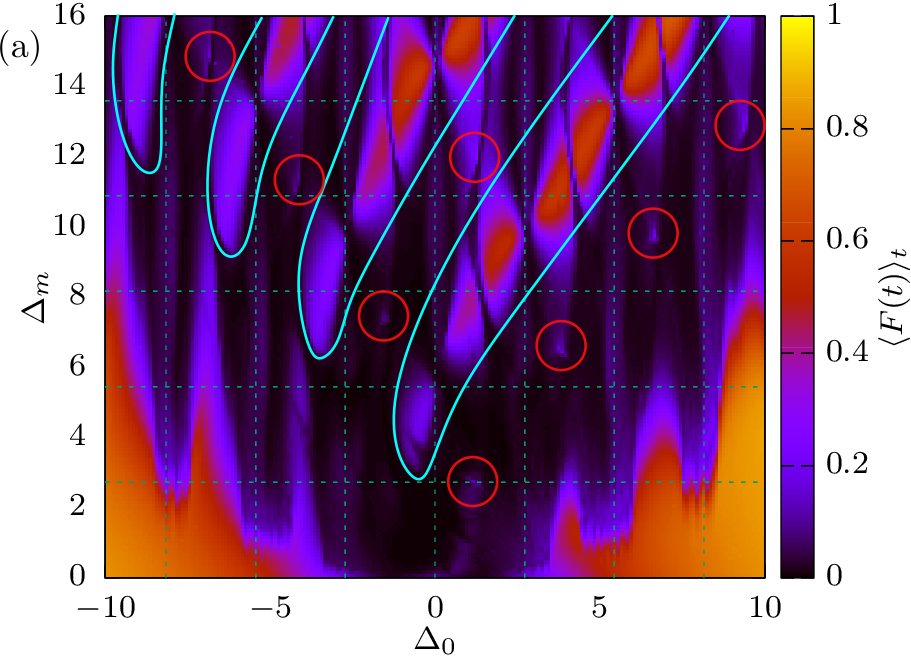}\\
 \includegraphics[width=0.43\textwidth]{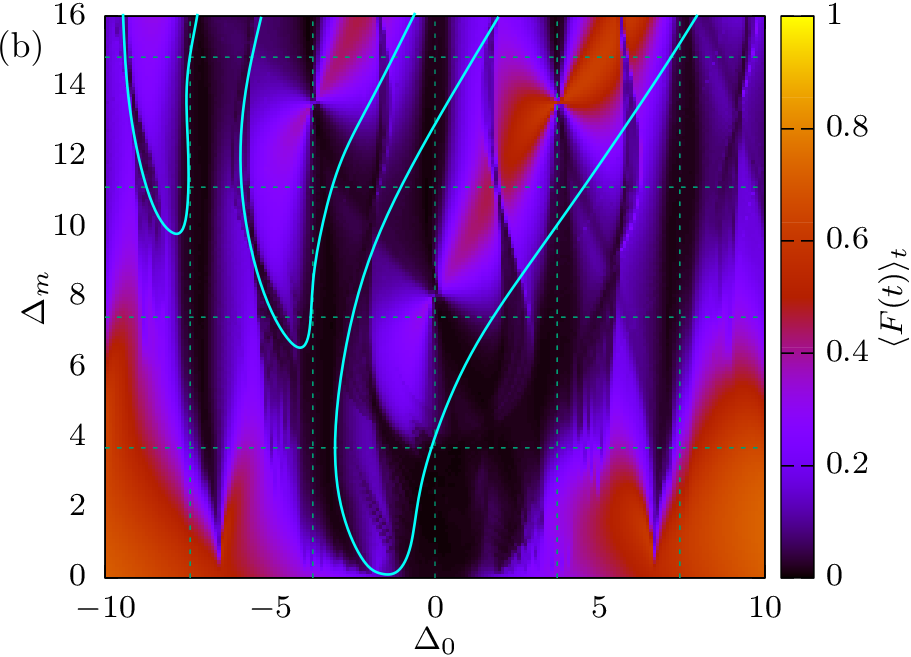}\\
\includegraphics[width=0.42\textwidth]{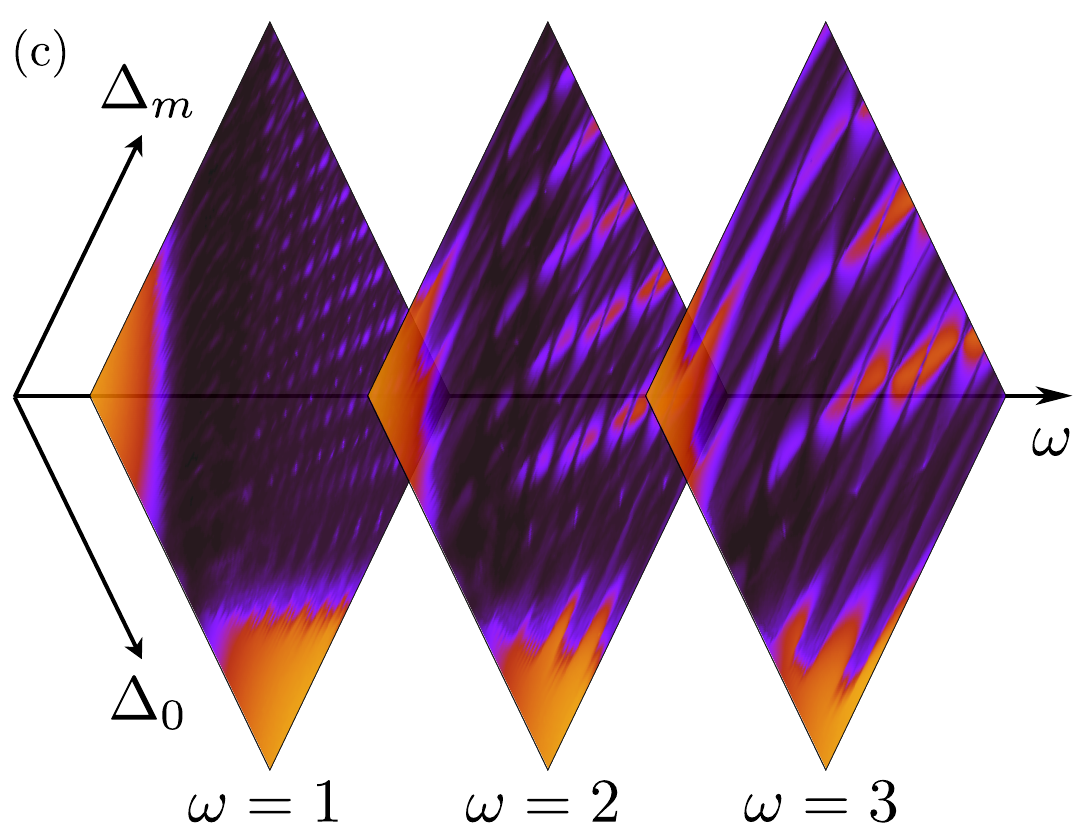}\\
 \caption{Average revival fidelity $\langle F(t)\rangle_t$ over time interval $[0, 100]$ for the PXP model with $\Omega=1$ and cosine drive at system size $N=20$. 
 The Type-1 peaks are encircled in red, while the regions containing Type-2 peaks are denoted by cyan lines.
 (a) N\'eel initial state at fixed frequency $\omega=2.72$. 
 (b) Polarized initial state at fixed frequency $\omega=3.71$.
 (c) N\'eel initial state for three different frequencies, $\omega\in\{1,2,3\}$, $\Delta_0\in[-10,10]$, $\Delta_m\in[0,16]$.
 }\label{fig:fidelity_scan_neel}
 \end{figure} 

In order to determine more precisely the optimal driving parameters, we scan the parameter space ($\Delta_0$, $\Delta_m$, $\omega$) for the highest average fidelity between the times $t=0$ and $t=\tau$,
\begin{eqnarray}\label{eq:cost}
\langle F\rangle_t = \frac{1}{\tau}\int_{0}^{\tau}F(t)\mathrm{d}t,
\end{eqnarray}
where we set $\tau=100$.
In Figs.~\ref{fig:fidelity_scan_neel}(a) and (b) we plot the average fidelity as a function of $\Delta_0$ and $\Delta_m$ at the fixed frequency $\omega$, for the N\'eel and polarized initial state. 
In this figure we restrict to the case $\Delta_m\geq0$ since $\langle F\rangle_t(\Delta_0, \Delta_m, \omega)=\langle F\rangle_t(-\Delta_0, -\Delta_m, \omega)$.
This symmetry comes from the fact that the dynamics of the system is exactly equal for $\pm\Delta(t)$. Changing the sign of $\Delta(t)$ only changes the sign of the quasienergy spectrum and multiplies the corresponding Floquet modes by the operator $\prod_j Z_j$ (with $Z_j$ being the Pauli $z$ matrix on site $j$), which commutes with the driving term and anticommutes with the PXP term.
The frequency is fixed to a value close to the revival frequency of the static model, i.e., we choose $\omega=2.72$ in (a) and $\omega=3.71$ in (b). These two figures have many similarities, such as parallel diagonal lines consisting of ``butterfly-shaped" peaks. However, a closer look reveals an important difference. The N\'eel state plot contains several additional peaks, which are very narrow and typically located between the bright diagonal lines. For example, one of these peaks is approximately around the point ($\Delta_0$,$\Delta_m$)=(1.15,2.67) and corresponds to an interesting optimal parameter regime for the N\'eel state, which will be studied below. We will call these narrow peaks ``Type-1" peaks. 

The other bright regions present in both Figs.~\ref{fig:fidelity_scan_neel}(a) and \ref{fig:fidelity_scan_neel}(b) will be dubbed ``Type-2 peaks'' (excluding $\lvert\Delta_0\rvert\gg1$ and $\Delta_m\lessapprox\lvert\Delta_0\rvert$). These parameter regions are broader, i.e., the parameters do not have to be finely tuned, and they typically occur at higher driving amplitudes than Type-1 peaks. Like Type-1, their position depends on the driving frequency, but the mechanism of revival stabilization is different. We will discuss Type-2 peaks below in relation to the polarized initial state.

It is important to note that the N\'eel and polarized state are two special cases. The cross sections in Figs.~\ref{fig:fidelity_scan_neel}(a) and (b) look very different for other initial product states, with low and very narrow peaks of the cost function instead of Type-2 peaks, while the Type-1 peaks are completely absent. One such example is shown in Fig.~\ref{fig:scan_random}(a) in Appendix~\ref{a:inhomogeneous}.

Finally,  Fig.~\ref{fig:fidelity_scan_neel}(c) shows three illustrative cross sections taken at different frequencies, $\omega \in \{1,2,3\}$. The initial state is the N\'eel state and the color scale 
corresponds to the value of the average fidelity cost function. We see that this cost function has many peaks (bright regions), which move and deform as $\omega$ is varied. The position of the fidelity peaks in ($\Delta_0$, $\Delta_m$) space depends approximately linearly on the driving frequency. In all three cases, there are two large bright regions in the $\lvert\Delta_0\rvert\gg1$ and $\Delta_m\lessapprox\lvert\Delta_0\rvert$ regime. When the static detuning $\Delta_0$ is large, the energy spectrum splits into disconnected bands. This means that the wave function cannot spread through the entire Hilbert space and the fidelity remains relatively high during the evolution, never dropping down to zero. These trivial regimes do not correspond to typical QMBS behavior, hence we exclude those  where the \emph{minimal} fidelity from $t=0$ to $t=100$ is non-zero. The optimal parameter regimes that will be discussed in detail below are summarized in Table~\ref{tab:parameters}.

\begin{table}
\begin{tabular}{ |c|c|c|c|c| }
\hline
Initial state & $\Delta_0$ & $\Delta_m$ & $\omega$ & Floquet spectrum \\
\hline
\multirow{1}{4em}{N\'eel} & 1.15 & 2.67 & 2.72 & two arcs\\ 
\hline
\multirow{2}{4em}{polarized} & 1.68 & -0.5 & 3.71 & one arc \\ 
& 0.64 & 7.55 & 2.90 & multiple arcs\\ 
\hline
\end{tabular}
\caption{Optimal driving parameters (cosine drive).}
\label{tab:parameters}
\end{table}

\begin{figure}[htb]
 \includegraphics[width=0.48\textwidth]{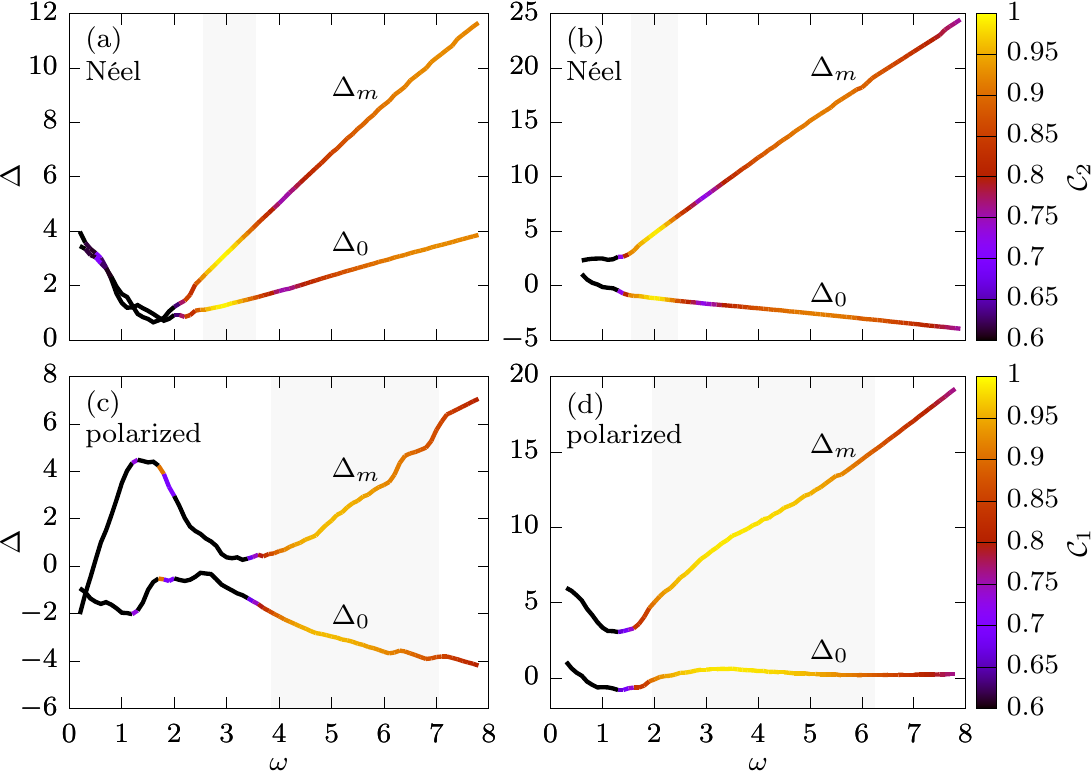}
 \caption{
 Optimal values of $\Delta_0$ and $\Delta_m$ depending on the driving frequency $\omega$. The color represents the value of $\mathcal{C}_k$ in Eq.~(\ref{eq:newcost}), with $k=2$ for the top panels (subharmonic response) and $k=1$ for the bottom panels (harmonic response).
 The shadowed regions mark the optimal frequency windows ($\mathcal{C}_k$ higher than 90\% of its maximal value).
 Top row: N\'eel state. (a) ``Main'' Type-1 peak (maximum around $\omega=2.72$). (b) Another Type-1 peak (maximum around $\omega=2$). Bottom row: Polarized state, Type-2 peaks. (c) Low amplitude. 
 (d) High amplitude. 
 }\label{fig:frequency_neel_pol}
 \end{figure}
 
Next we determine the optimal window of driving frequencies. In order to do this, we choose one of the narrow peaks of the cost function in Fig.~\ref{fig:fidelity_scan_neel} and follow it as we vary the frequency $\omega$. The driving parameters $\Delta_0$ and $\Delta_m$ are reoptimized for each frequency. When we leave the optimal frequency window, the peak disappears, but other peaks may become more prominent, as summarized in Fig.~\ref{fig:frequency_neel_pol}.
We note that even though the original peak is no longer clearly visible, there is still a local maximum of the cost function and we are able to track it outside of the optimal frequency window.
In order to characterize  the quality of revivals, in Fig.~\ref{fig:frequency_neel_pol} we used the following cost function
\begin{eqnarray}\label{eq:newcost}
\mathcal{C}_l = \frac{1}{M}\sum_{n=1}^{M}\left[F(l n T)-F(l(n-1/2)T)\right],
\end{eqnarray}
where $T$ is the driving period and $M$ denotes the number of revivals averaged over. The integer $l$ defines the order of the response, e.g., $l=1$ for the harmonic and $l=2$ for the subharmonic responses we will study below. In both cases we average over $lM=40$ driving periods. 
The value of $\mathcal{C}_l$ corresponds to the difference between average maximum fidelity and average minimum fidelity over the first $M$ revivals, with the value of 1 corresponding to perfect revivals. In Fig.~\ref{fig:frequency_neel_pol} we use $\mathcal{C}_l$ instead of the previously introduced average fidelity because it gives better results when comparing different frequencies. Namely, as the driving frequency is increased, the revivals become more dense and the width of an individual revival increases in comparison with the revival period, which then results in increased average fidelity. Therefore, decaying revivals at higher frequencies can lead to higher average fidelity than stabilized and constantly high revivals at lower frequencies. The new cost function does not suffer from this problem and can accurately estimate the quality of revivals independently of the frequency. However, we still use the first cost function for fixed frequency scans, such as those in Fig.~\ref{fig:fidelity_scan_neel}, since the average fidelity can capture all types of revivals (e.g., harmonic, subharmonic or even incommensurate with the driving frequency), without the need to specify their expected frequency.

In Fig.~\ref{fig:frequency_neel_pol}(a) we plot the data for the peak that passes through a set of N\'eel state optimal parameters ($\Delta_0=1.15$, $\Delta_m=2.67$, $\omega=2.72$). We look at $M=20$ revivals in evaluating the cost function $\mathcal{C}_2$ in Eq.~(\ref{eq:newcost}). The dependence of $\Delta_0$ and $\Delta_m$ on $\omega$ is seen to be approximately linear.  In Fig.~\ref{fig:frequency_neel_pol}(a) we see that there is indeed a window of optimal frequencies, approximately $2.5<\omega<3.5$, where we can obtain a good subharmonic response to periodic driving. This window is centered at a frequency slightly higher than twice the bare revival frequency of the N\'eel state in the pure PXP model ($\omega_0\approx2.66$). In Appendix~\ref{a:scaling}, we study the system size scaling of this optimal driving frequency, see Fig.~\ref{fig:frequency_neel_size}. For $N \geq 10$, the optimal frequency window for the N\'eel state barely changes with increasing system size. Note that there is a minimal frequency in Fig.~\ref{fig:frequency_neel_pol}(a) below which it is impossible to obtain revivals. On the other hand, it is possible to enhance the revivals in the high frequency regime, but the lifetime of these revivals is finite and decreases with increasing frequency.

 We can repeat the same procedure for other Type-1 peaks, see Fig.~\ref{fig:frequency_neel_pol}(b) for one example. In this case we start from the optimal parameters that correspond to the ``one-loop" trajectory, which will be studied in more detail in Sec.~\ref{ss:trajectory} ($\Delta_0=-1.17$, $\Delta_m=4.81$, $\omega=2.00$).
 The dependence of optimal $\Delta_0$ and $\Delta_m$ on $\omega$ is again linear, but the optimal frequency window is now more narrow and centered around $\omega=2$, which significantly deviates from $\omega_0\approx2.66$. The results for other Type-1 peaks are found to be similar (data not shown).

\begin{figure*}[tbh]
 \includegraphics[width=0.32\textwidth]{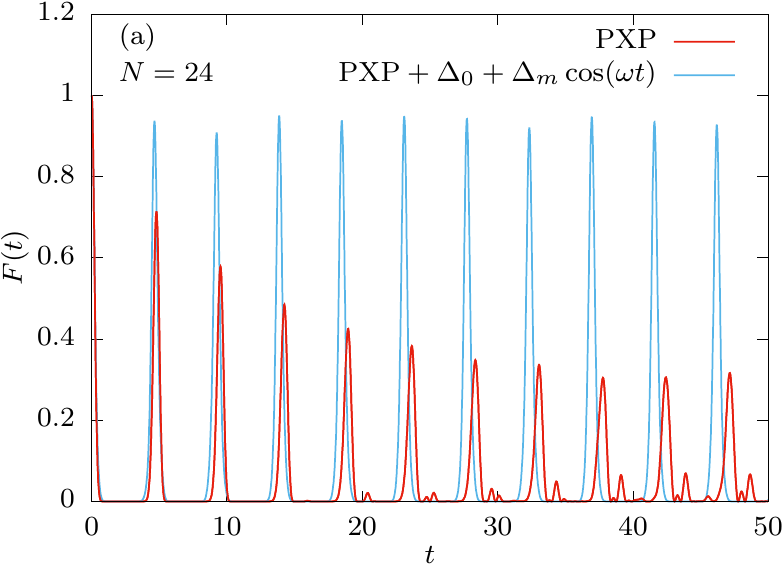}
 \includegraphics[width=0.32\textwidth]{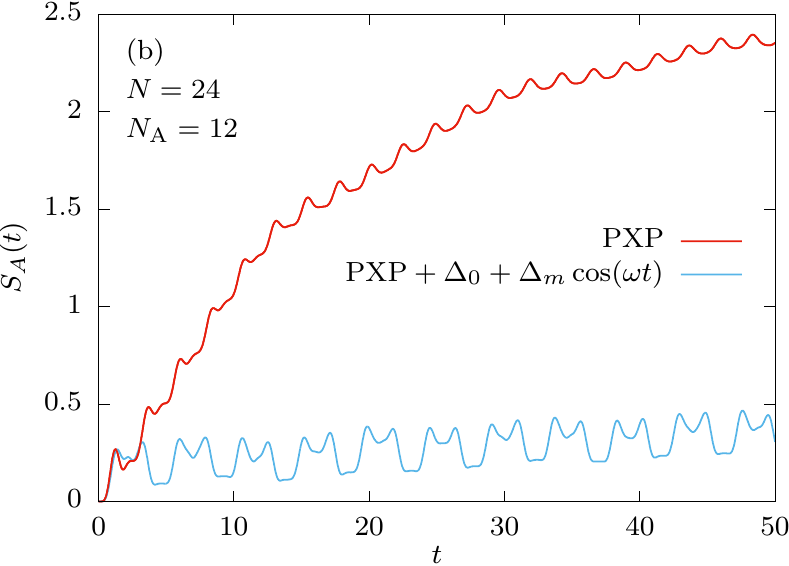}
 \includegraphics[width=0.31\textwidth]{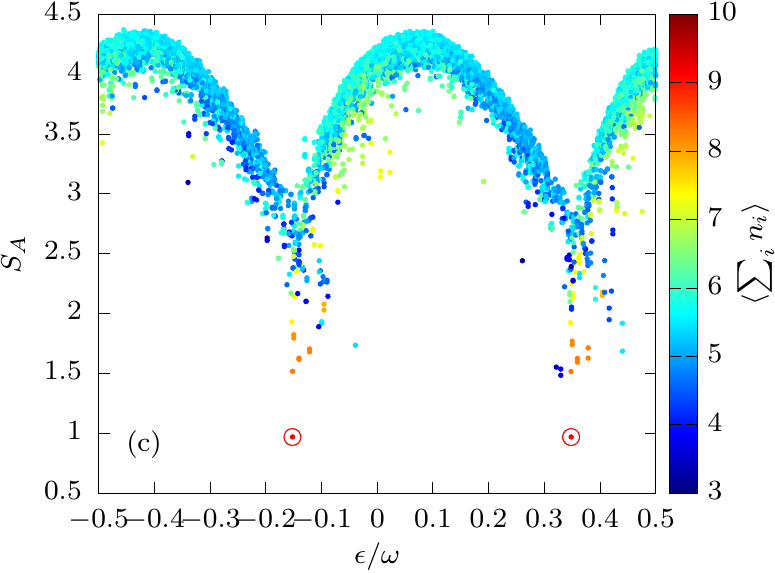}
 \caption{Driven N\'eel state ($\Delta_0=1.15$, $\Delta_m=2.67$, $\omega=2.72$). 
 (a) Quantum fidelity $F(t)$ for system size $N=24$. (b) Entanglement entropy $S_A$ for a subsystem with $N_A=12$ sites. (c) Entanglement entropy of the Floquet modes for system size $N=20$. Color scale shows the average number of  excitations, $\langle \sum_i n_i \rangle$, in each Floquet mode. 
 }\label{fig:driven_neel}
 \end{figure*}

Finally, this procedure can also be adapted to the harmonic response, e.g. for the polarized and other initial states with the cost function $\mathcal{C}_1$. The results for the polarized state and $M=40$ in the low- and high-amplitude regime (parameters from Table~\ref{tab:parameters}) are shown in Fig.~\ref{fig:frequency_neel_pol}(c) and (d), respectively. In both regimes, the optimal frequency window is much wider than for the N\'eel state. 
 
\section{N\'eel and polarized state revivals}\label{s:revivals}

In this Section we investigate in more detail the optimally driven N\'eel and polarized initial states. 

\subsection{N\'eel state}

Using the procedure described in Sec.~\ref{s:cosine} -- scanning the parameter space for the highest average fidelity -- we found the following optimal parameters for the N\'eel state: $\Delta_0=1.15$, $\Delta_m=2.67$, $\omega=2.72$. 
This is the lowest-amplitude Type-1 peak in Fig.~\ref{fig:fidelity_scan_neel}(a). 
In addition to brute force scanning, we also employed the simulated annealing optimization algorithm, which produced similar optimal values. We note that our driving parameters slightly differ from those considered in Ref.~\onlinecite{Bluvstein2021},  $\Delta_0=1.00$ and $\Delta_m=2.00$, which were obtained at the fixed frequency $\omega=2.66$. 

Fig.~\ref{fig:driven_neel}(a) shows the evolution of quantum fidelity for the optimal parameters quoted above. In  contrast to the undriven case, the revivals are clearly enhanced and remain stable over long times. Note that the revival frequency is half the driving frequency. This is an example of period doubling (subharmonic response), which was related to a discrete time crystal in Ref.~\onlinecite{Bluvstein2021}. The growth of the bipartite von Neumann entanglement entropy $S_A$ is strongly suppressed by driving, as shown in Fig.~\ref{fig:driven_neel}(b). To evaluate  $S_A$, we partition the chain into two equal halves and evaluate the reduced density matrix for one of the subsystems, $\rho_A(t)=\mathrm{tr}_B \ket{\psi(t)}\bra{\psi(t)}$, by tracing out the other ($B$) subsystem. This yields $S_A = - \mathrm{tr} \; \rho_A \ln\rho_A$.

Furthermore, we numerically construct the one period evolution operator of the optimally driven system and diagonalize it in order to obtain the corresponding Floquet modes. The bipartite entanglement entropy versus the quasienergy of all the Floquet modes is plotted in Fig.~\ref{fig:driven_neel}(c). The color represents the expectation value of the number of excitations $\langle\sum_i n_i\rangle$ for each mode. This entropy plot has an interesting ``two arc'' structure. The origin of these two arcs will be explained in Sec.~\ref{ss:arcs} below. The two lowest-entropy modes (marked by the red circles) also have the highest overlap with the N\'eel state. These two modes are said to be $\pi$-paired, since they are separated by $\Delta\epsilon\approx\omega/2=\pi/T$ in quasienergy. Such two $\pi$-paired Floquet modes were also observed in Ref.~\onlinecite{Maskara2021}, where the cosine drive was approximated by a delta function pulsed drive, as well as in other systems, which exhibit time-crystalline behavior but no scarring \cite{Pizzi2020}. We find that the two modes are approximately $(\lvert \mathbb{Z}_2\rangle+\lvert \mathbb{Z}'_2\rangle)/\sqrt{2}$ and $(\lvert \mathbb{Z}_2\rangle-\lvert \mathbb{Z}'_2\rangle)/\sqrt{2}$, which explains the observed period doubling~\cite{Su2022}. However, the subharmonic response becomes harmonic if we work in the momentum $k=0$ sector and choose a translation invariant initial state $(\lvert \mathbb{Z}_2\rangle+\lvert \mathbb{Z}'_2\rangle)/\sqrt{2}$. In that case, there will still be two arcs in the entropy plot, but only one high-overlap Floquet mode.

In addition to $(\Delta_0,\Delta_m)=(1.15,2.67)$ parameters, there are several other Type-1 peaks in the cost-function scan at fixed frequency $\omega=2.72$ that we have seen in Fig.~\ref{fig:fidelity_scan_neel}(a).
The entanglement entropy plots of their Floquet modes also display the distinct ``two arcs'' structure.
These other Type-1 peaks also lead to enhanced revivals, although with a finite lifetime at this particular driving frequency. However, some of those peaks are higher at other frequencies and can lead to completely stabilized revivals, see for example Fig.~\ref{fig:frequency_neel_pol}(b) and the discussion in Sec.~\ref{ss:trajectory}. 
The number, position and height of Type-1 peaks depends on the driving frequency, while the driving amplitude of peaks that result in stable revivals is typically lower than that of Type-2 peaks.

\subsection{Polarized state}

For the polarized initial state, particularly in the low-amplitude driving regime, the optimal frequency value is found to be more sensitive to the cost function ($\langle F\rangle_t$ or $\mathcal{C}_1$) used for the optimization.  In particular, the $\mathcal{C}_1$ cost function results in a larger value of the driving amplitude, see Fig.~\ref{fig:frequency_neel_pol}(c). Below we will use a set of parameters close to the low-amplitude edge of the optimal frequency window. These parameters were also used to drive the polarized state in a recent experiment on the Bose-Hubbard quantum simulator of the PXP model \cite{Su2022}.

\begin{figure*}[bth]
 \includegraphics[width=0.32\textwidth]{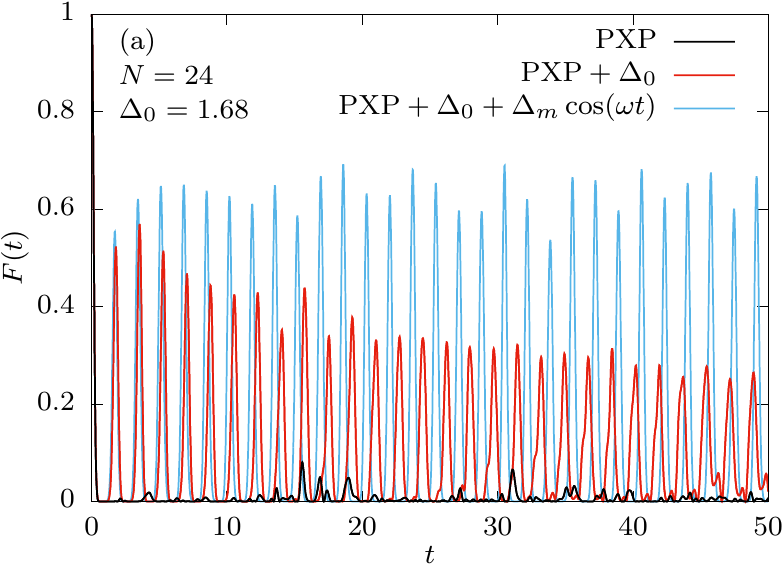}
 \includegraphics[width=0.32\textwidth]{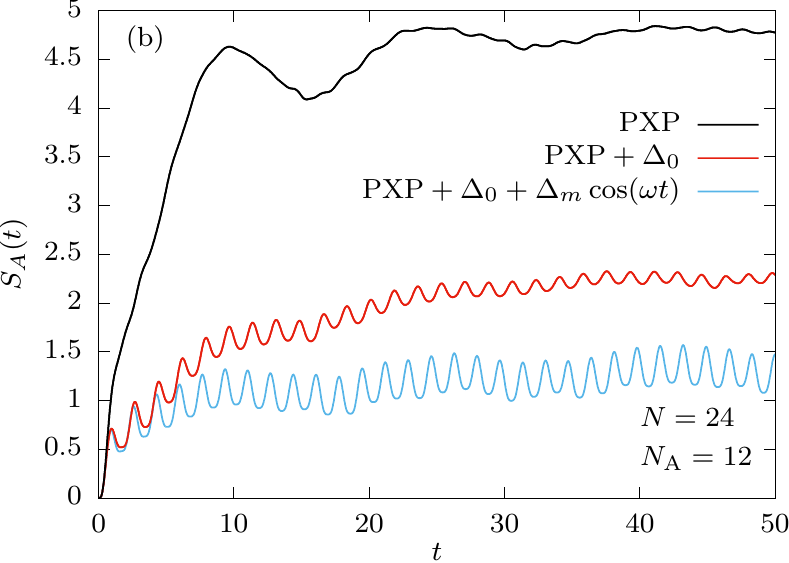}
 \includegraphics[width=0.31\textwidth]{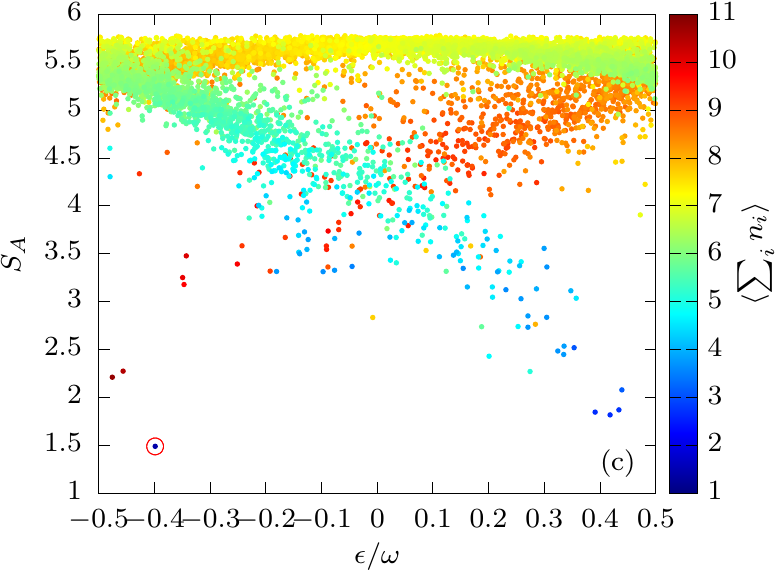}\\
 \includegraphics[width=0.32\textwidth]{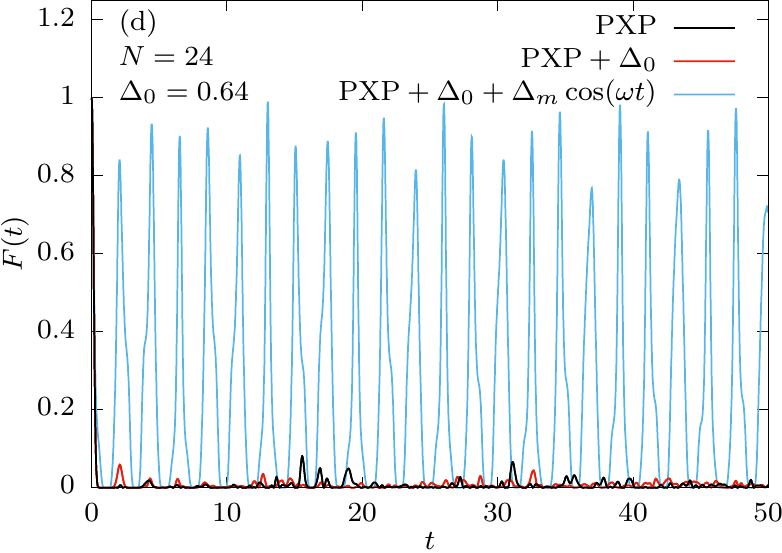}
 \includegraphics[width=0.32\textwidth]{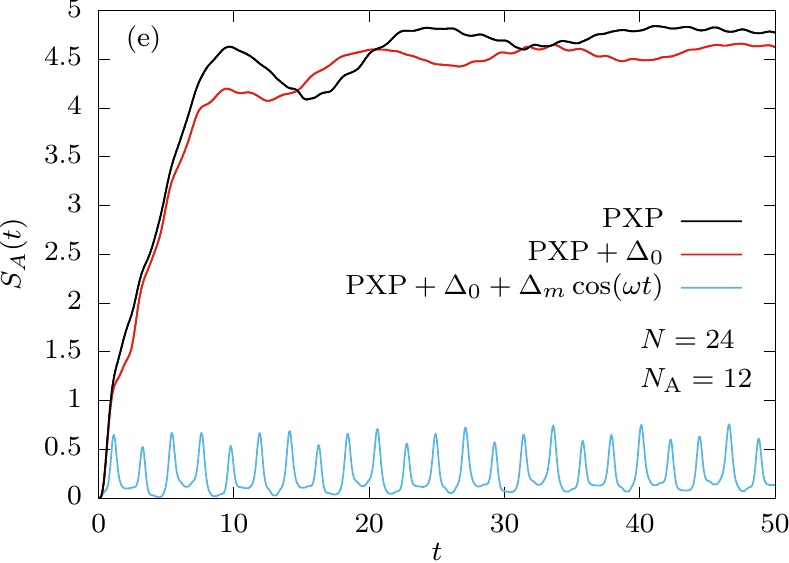}
 \includegraphics[width=0.31\textwidth]{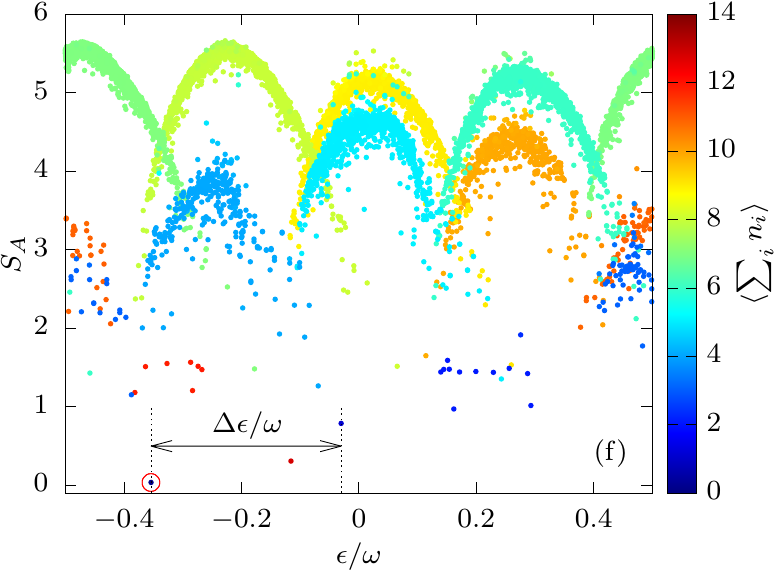}
 \caption{Driven polarized state. Top row: ``Low amplitude" driving regime ($\Delta_0=1.68$, $\Delta_m=-0.50$, $\omega=3.71$). 
 (a) Quantum fidelity, system size $N=24$. (b) Entanglement entropy for the subsystem size $N_A=12$. (c) Entanglement entropy of Floquet modes, $N=26$ in zero momentum sector. Color scale shows the average number of  excitations, $\langle \sum_i n_i \rangle$, in each Floquet mode.
 Bottom row: Same as the top row, but for ``high amplitude" driving regime ($\Delta_0=0.64$, $\Delta_m=7.55$, $\omega=2.90$). 
 }\label{fig:driven_polarized}
 \end{figure*}

Although the fully polarized state $\lvert 0\rangle$ shows no QMBS behavior and quickly thermalizes in the pure PXP model, it was shown that this can be changed by applying a static detuning term and then further enhanced by periodic driving~\cite{Su2022}. Figure~\ref{fig:driven_polarized}(a) shows the evolution of quantum fidelity for the polarized state in the pure PXP model (black), with static detuning (red), and with periodic driving (blue). Similar to the N\'eel state, the growth of entanglement entropy is strongly suppressed by the static detuning and even more so with periodic driving, see Fig.~\ref{fig:driven_polarized}(b). The entanglement entropies and the numbers of excitations for all Floquet modes of the optimally driven system ($\Delta_0=1.68$, $\Delta_m=-0.50$, $\omega=3.71$) can be observed in Fig.~\ref{fig:driven_polarized}(c). This driving frequency approximately matches that of revivals in a static system with detuning $\Delta_0=1.68$ [red line in Fig.~\ref{fig:driven_polarized}(a)], where QMBS for the polarized state has been previously observed~\cite{Su2022}. 
However, unlike the N\'eel state, there is now only one arc and a single Floquet mode that has high overlap with the polarized state ($\approx64\%$, encircled in red).

Apart from this set of driving parameters, which were studied in Ref.~\onlinecite{Su2022}, there are also other sets of driving parameters that lead to robust revivals from the polarized state. One such example is shown in Figs.~\ref{fig:driven_polarized}(d)-\ref{fig:driven_polarized}(f) ($\Delta_0=0.64$, $\Delta_m=7.55$, $\omega=2.90$). Upon first glance, this situation is completely different from Figs.~\ref{fig:driven_polarized}(a)-\ref{fig:driven_polarized}(c). In the latter case, the revivals essentially stem from static detuning, which creates scarring, and the driving further stabilizes them. By contrast, in the former case, the revivals are generated by periodic driving. The optimal driving frequency was found using simulated annealing and it does not appear to correspond to any special frequency in the static system. Moreover, the Floquet mode entanglement entropy plot also has a different structure, as can be observed in Fig.~\ref{fig:driven_polarized}(e). There are now multiple arcs ($N/2+1$ of them), each with a well defined excitation number. This suggests that the Floquet Hamiltonian is fragmented in this parameter regime (see further discussion in Sec.~\ref{ss:arcs} below). There is a single Floquet mode that has near-unity overlap ($>90\%$) with the polarized state, which accounts for the revivals. The mode belongs to its own one-dimensional excitation number sector, since the  polarized state is the only state that contains no excitations. The mode is furthermore separated from the neighboring $\langle\sum_i n_i\rangle=1$ sector by a gap $\Delta\epsilon$, which practically does not depend on the system size -- see Fig.~\ref{fig:size_scaling_gap_pol} in Appendix~\ref{a:scaling}. 

Based on these insights, in Appendix~\ref{a:inhomogeneous} we present a modified, spatially inhomogeneous cosine driving scheme, which can enhance the revivals from any initial product state, including generic states that do not display QMBS. The main idea is to change the driving protocol in such a way that the desired initial state becomes the ground state of the new time-dependent term. Driving with optimal parameters then results in similar phenomenology to the polarized state in the high-amplitude regime (Fig.~\ref{fig:driven_polarized}(d)-(f)), including the fragmentation of the Floquet Hamiltonian into sectors with different quantum numbers.

We note that both of the driving regimes presented in Fig.~\ref{fig:driven_polarized} belong to what we referred to as Type-2 peaks.  In these cases, the optimal parameters do not require particular fine-tuning, i.e., there is a fairly broad range of parameter values for which the driving stabilizes QMBS revivals, unlike the Type-1 peaks where such regions are more narrow and finding optimal parameters requires more careful optimization. It was not possible to produce a subharmonic response for the polarized state, which is expected since it does not have a translated partner state, such as $\mathbb{Z}_2$ and $\mathbb{Z}'_2$. Since the two N\'eel states also belong to their own excitation number sector in Fig.~\ref{fig:driven_polarized}(f), it is also possible to use the $\Delta_0=0.64$, $\Delta_m=7.55$, $\omega=2.90$ parameters to stabilize the revivals from these states, as will be explained in Sec.~\ref{s:scars}.

\subsection{Origin of ``multiple arcs'' structure}\label{ss:arcs}

In order to explain the structure of entanglement entropy plots in Figs.~\ref{fig:driven_neel} and \ref{fig:driven_polarized}, we treat the PXP term in Eq.~\eqref{eq:ht} as a perturbation. First we consider the $\Omega=0$ case when the Hamiltonian is diagonal but still time periodic. The Floquet modes are then product states and their quasienergies depend only on the value of static detuning $\Delta_0$ and their number of excitations $\langle\sum_in_i\rangle$, while the quasienergy spectrum has periodicity $\omega$,
\begin{eqnarray}\label{eq:quasienergies}
\epsilon=\langle\sum_in_i\rangle\Delta_0\mod\omega.
\end{eqnarray}
As the number of excitations increases, the Floquet modes are winding around the spectrum. Different ratios of $\Delta_0/\omega$ will result in different structures. When the PXP term is turned on, these states start to hybridize. 
However, only those states that are close in excitation number can hybridize, since the PXP term can only change this number by $\pm 1$. They also need to be close in quasienergy, but what is ``close enough" depends on the magnitude of the PXP term in the Floquet Hamiltonian. Since the cosine drive is difficult to treat analytically, especially for the N\'eel state optimal parameters, which are in the non-perturbative regime, in the following Sec.~\ref{s:square} we introduce a more tractable square pulse approximation of the cosine driving protocol.

In the parameter regime optimal for the N\'eel state (Type-1), the $\Omega=0$ Floquet modes form two diagonal lines (even and odd excitation numbers). Only the states belonging to the same diagonal will mix when $\Omega>0$, which will result in two separate structures at $\Omega=1$. This is the origin of the two arcs in Fig.~\ref{fig:driven_neel}(c). Note that the Floquet modes in the two arcs no longer consist of the states with only even or only odd number of excitations. No symmetry that distinguishes between the two arcs could be identified.

The situation is different for the polarized state high-amplitude optimal parameters (Type-2). At $\Omega=0$, the Floquet modes that are close in quasienergy are far apart in excitation number, and the magnitude of the PXP term in the Floquet Hamiltonian is small for these parameters, so there is no mixing between the sectors even when $\Omega=1$. This results in $N/2+1$ separate sectors as shown in Fig.~\ref{fig:driven_polarized}(f), each containing only the modes with a certain number of excitations. The low-amplitude regime from Fig.~\ref{fig:driven_polarized}(c) is a special case where the smaller arcs merge into one large arc, while the Floquet modes are still arranged according to their numbers of excitations and the $\langle\sum_i n_i\rangle=0$ sector is sufficiently separated from the rest.

\section{Square pulse drive}\label{s:square}

The cosine drive in Eq.~\eqref{eq:cos_drive} is difficult to treat analytically. Therefore, it is desirable to find a simpler model that accurately captures the dynamics, at least at the qualitative level. Previous works\cite{Bluvstein2021,Maskara2021} have considered the following delta function pulse driving model:
\begin{eqnarray}\label{eq:deta}
\Delta(t)&=&\theta\sum_{n\in\mathbb{Z}}\delta(t-nT),
\end{eqnarray}
where $T=2\pi/\omega$ is the driving period and $\theta$ is a parameter of the drive. This driving protocol results in perfect revivals from any initial state when $\theta=\pi$ due to a special symmetry property of the Floquet operator in that case~\cite{Maskara2021}. However, the N\'eel state is seen to be special when $\theta$ is varied away from $\pi$: while the revival from other initial states sharply decays, the N\'eel revival stays robust over a finite range of $\theta$ in the vicinity of $\pi$. Although this model has been well understood, its relation to the cosine driving scheme is not completely transparent.

 \begin{figure}[t]
 \includegraphics[width=0.44\textwidth]{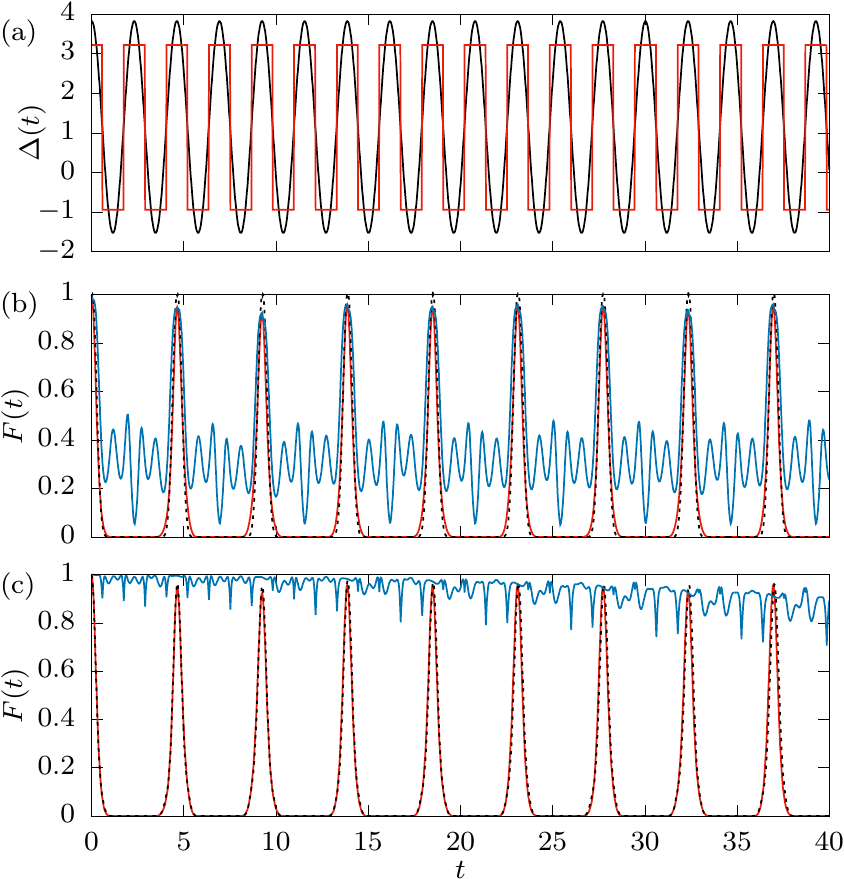}\\
 \includegraphics[width=0.44\textwidth]{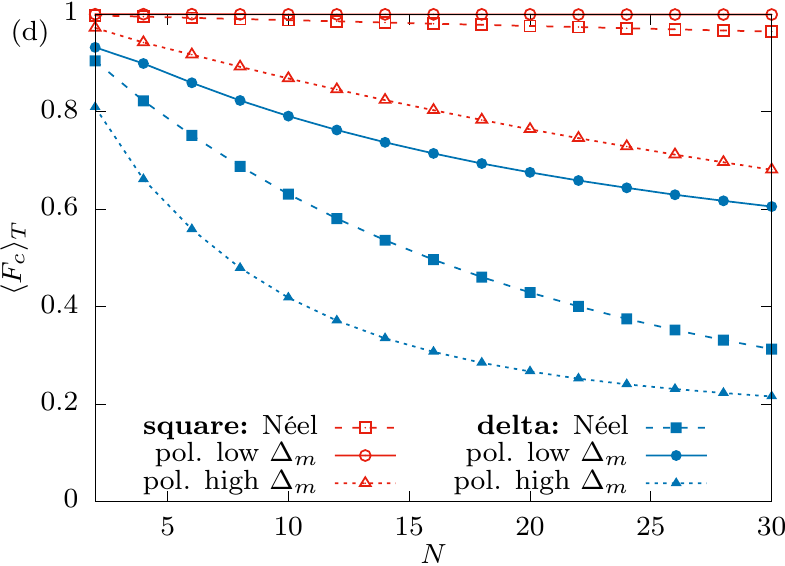}
 \caption{(a) Cosine drive (black) vs square pulse drive (red).
 (b) Red: Cosine drive fidelity, $F(t)=\lvert\langle\psi(0)\lvert\psi(t)\rangle\lvert^2$. Black (dashed): Delta function pulse fidelity, $F'(t)=\lvert\langle\psi(0)\lvert\psi'(t)\rangle\lvert^2$. Blue: Comoving fidelity $F_c(t)=\lvert\langle\psi(t)\lvert\psi'(t)\rangle\lvert^2$. N\'eel state, $L=20$, optimal driving parameters. (c) Same for square pulse instead of delta function pulse.
 (d) System size scaling of comoving fidelity averaged over the first driving period for the square (red) and delta function pulse (blue). Optimal driving parameters: N\'eel state (squares), polarized state low-amplitude (circles), polarized state high-amplitude (triangles). 
 }\label{fig:square_pulse}
 \end{figure} 

We will now show that a simple square pulse model is a better approximation for the cosine drive. The square pulse protocol is defined by
\begin{eqnarray}\label{eq:square}
\Delta(t)&=&\Delta_0+\Delta_m(t),\nonumber\\
\Delta_m(t)&=&\begin{cases}
              \Delta_m, &0\leq t\leq T/4\\
              -\Delta_m, &T/4< t\leq 3T/4\\
              \Delta_m, &3T/4< t\leq T,
              \end{cases}
\end{eqnarray}
as illustrated in Fig.~\ref{fig:square_pulse}(a).
We note that similar driving schemes were previously considered in Refs.~\onlinecite{Mukherjee2020a,Mukherjee2020b}, however our square pulse model differs from those by a quarter of driving period time shift, which results in very different Floquet dynamics.
 
A comparison of the delta pulse in Eq.~\eqref{eq:deta} and the square pulse in Eq.~\eqref{eq:square} is shown in  Fig.~\ref{fig:square_pulse}(b) and (c), where we plot the quantum fidelity $F(t)$ for the exact evolution under cosine drive, as well as the fidelity for the system driven by the delta or square pulse.  Moreover, we directly evaluate and plot the comoving fidelity, $F_c(t)=\lvert\langle\psi(t)\lvert\psi'(t)\rangle\lvert^2$, which represents the overlap between the wave function evolved using the cosine driving protocol, $\lvert\psi(t)\rangle$, and the one evolved using the approximate protocol, $\lvert\psi'(t)\rangle$ (delta or square pulse).
 Here, the initial state is the N\'eel state and the driving parameters are set to their optimal values ($\Delta_0=1.14$, $\Delta_m=2.08$, $\omega=2.72$ for the square pulse). We note that the optimal driving frequency and static detuning are approximately the same for both the cosine and square pulse drive, but the driving amplitude has to be reoptimized and is approximately $20\%$ lower in the latter case.  
 For the delta function pulse, we use $\theta=\pi$ and the same frequency as the cosine drive.
 
 The comoving fidelity is seen to be much higher for the square pulse drive, compare Figs.~\ref{fig:square_pulse}(b) and (c). This signals that the square pulse driving protocol is a much better approximation for the cosine drive than the delta function pulse. In particular, while the delta pulse is highly accurate near the revival points, the square pulse remains accurate throughout the evolution, including away from the revival points. This difference becomes more pronounced in larger system sizes, as can be observed in Fig.~\ref{fig:square_pulse}(d),
 where we plot the scaling of the comoving fidelity, averaged over the first driving period with the system size $N$.
In general, the agreement between the cosine drive and square pulse is better when the driving frequency is high and amplitude low. 
For example, the agreement is particularly good for the polarized state in the low-amplitude regime, where the comoving fidelity is almost equal to $1$ over many driving periods. 
In addition to accurately capturing the dynamics of the cosine drive, the square pulse drive also reproduces the two special Floquet modes and a ``two arc'' entanglement entropy spectrum resembling the one in Fig.~\ref{fig:driven_neel}(c).
Finally, knowing the optimal driving parameters for the square pulse allows us to estimate the parameters for the cosine drive by simply using the same driving frequency and static detuning while increasing the amplitude by $25\%$.

The optimal driving parameters identified for the N\'eel state ($\Delta_0=1.15$, $\Delta_m=2.67$, $\omega=2.72$) and for the polarized state ($\Delta_0=1.68$, $\Delta_m=-0.50$, $\omega=3.71$) are all of the same order of magnitude as $\Omega=1$, the characteristic energy scale of the static Hamiltonian. Thus, the stabilization of revivals in these cases due to periodic driving appears to be a non-perturbative effect. Nevertheless, for the square pulse protocol, we can obtain useful analytical insights into the mechanism of the revival stabilization  in two extreme limits: the limit of high amplitudes and the limit of high driving frequency. In the remainder of this section, we discuss separately these two cases.

\subsection{High-amplitude regime}

When the driving amplitude $\Delta_m$ is high, the square pulse allows us to describe the dynamics perturbatively.  At first order in the small parameter $\lambda_{\pm} \equiv \Omega/(\Delta_m \pm \Delta_0)\ll 1 $, it straightforward to obtain the Floquet Hamiltonian using the Pauli algebra. We start from the Hamiltonian
\begin{eqnarray}
H(t) = -\frac{\Delta(t)}{2}\sum_{j}(I_j+Z_j)-\Omega\sum_{j}\tilde{X}_j,
\end{eqnarray}
where $\tilde{X}_j \equiv P_{j-1}X_j P_{j+1}$ denotes the Pauli $x$ matrix dressed by projectors, and $\Delta(t)$ is given by the square pulse in Eq.~(\ref{eq:square}).  We ignore the energy shift $-\Delta(t)/2$ and assume $\Delta_m \pm\Delta_0\gg\Omega>0$. 

 \begin{figure}[tbh]
 \includegraphics[width=0.3\textwidth]{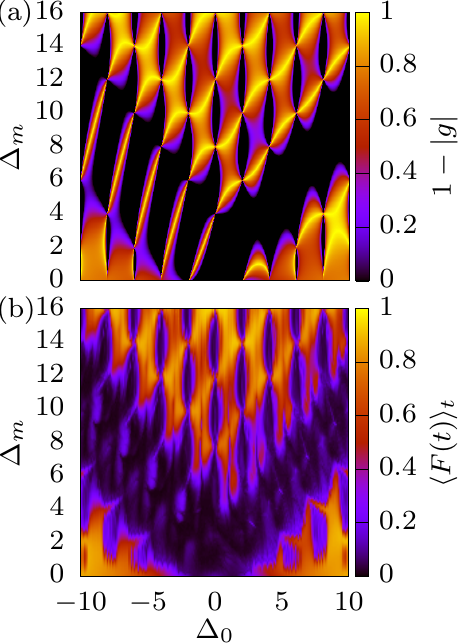}
 \caption{Analytical versus numerical results. N\'eel state, $\omega=2$. (a) $1-\lvert g_{T,\Delta_0,\Delta_m}\rvert$. Note that some values are outside of $[0,1]$ interval. (b) Average fidelity $\langle F(t)\rangle_t$.
 }\label{fig:floq_ham}
 \end{figure} 

The evolution operator from $t=0$ to $t=T/4$ and from $t=3T/4$ to $T$ is denoted $U_+$ and from $t=T/4$ to $t=3T/4$ by $U_-$. At first order $\lambda_{\pm}\ll 1 $, we can factorize these evolution operators 
\begin{eqnarray}\label{eq:U_plus_minus}
U_+ &\approx& \prod_{j}e^{i\left(\lambda_+ \tilde{X}_j+ Z_j\right)\frac{T}{4}},
\\
U_-&\approx&\prod_{j}e^{i\left(\lambda_{-}  \tilde{X}_j+Z_j\right)\frac{T}{2}},
\end{eqnarray}
since the commutator between $\lambda \tilde{X}_j + Z_j$ and $\lambda \tilde{X}_{j'} + Z_{j'}$ on different sites $j$ and $j'$ is second order in $\lambda$. We can then expand the exponentials using the properties of Pauli operators ($Z^2=I$ as well as $\tilde X^2 = I$) in terms of sine and cosine functions, and evaluate the total evolution operator for one period $U=U_+U_-U_+$. At first order in $\lambda$, the final expression can be easily converted back into exponential form, from which the first-order Floquet Hamiltonian can be directly read off. While the algebra is tedious, the calculation is straightforward and results in the following first-order Floquet Hamiltonian
\begin{eqnarray}\label{eq:floq_ham}
\nonumber H_F^{(1)}&=& -\frac{\Delta_0}{2} \sum_i g_{T,\Delta_0,\Delta_m}  P_{i-1}X_iP_{i+1}+Z_i, \\
\nonumber g_{T,\Delta_0,\Delta_m} &=& \frac{2\Omega}{\Delta_m^2-\Delta_0^2}\left\{\left[1+2\frac{\sin\frac{(\Delta_m-\Delta_0)T}{4}}{\sin\frac{\Delta_0 T}{2}}\right]\Delta_m-\Delta_0 \right\}.\\
\end{eqnarray}
The first-order Floquet Hamiltonian contains a PXP term whose magnitude is controlled by the prefactor $g$, which depends on the values of driving parameters $\Delta_0$, $\Delta_m$ and the drive period  $T=2\pi/\omega$. 

The $g$ prefactor explains several features of the fidelity phase diagram, as can be seen in Fig.~\ref{fig:floq_ham}. When $g$ is small, the Floquet Hamiltonian is almost diagonal and the Floquet modes are close to product states, which have well defined numbers of excitations. As long as $g$ is small enough [bright regions in Fig.~\ref{fig:floq_ham}(a)], there is no mixing between different excitation-number sectors. This leads to fragmentation and the polarized state ends up in its own one-dimensional sector, while the two N\'eel states belong to another two-dimensional sector. These states are therefore close to a Floquet mode and keep reviving in the driven system, which leads to high average fidelity [bright regions in Fig.~\ref{fig:floq_ham}(b)]. This explains the origin of Type-2 peaks.

Conversely, when $g$ is large (for example around the diagonals $\Delta_0=\pm\Delta_m$) the sectors are mixed and the Floquet modes have large variance of the excitation number, meaning that they are superpositions of many product states with different excitation numbers. We do not expect revivals in these regions and Fig.~\ref{fig:floq_ham}(b) confirms that this is mostly the case. However, the optimal driving parameters for the N\'eel state (Type-1 peaks) are actually in the dark region along the diagonal.

Some features of the phase diagram in Fig.~\ref{fig:floq_ham}(b) cannot be explained only by Eq.~\eqref{eq:floq_ham}. Irrespective of the value of $g$, there are dark lines (no revivals) around $\Delta_0=n\omega/2$, where $n$ is an integer (this is more visible for the cosine drive in Fig.~\ref{fig:fidelity_scan_neel}). The explanation is related to the quasienergies of the Floquet modes when the PXP term is turned off ($\Omega=0$), see the discussion in Sec. \ref{ss:arcs}. In this case, the Hamiltonian is diagonal and the Floquet modes are product states. Their quasienergies depend only on the value of static detuning $\Delta_0$ and the Floquet spectrum has periodicity $\omega$, see Eq.~\eqref{eq:quasienergies}.
When $\Delta_0=n\omega$, all $\Omega=0$ modes have the same quasienergy $\epsilon=0$. For $\Omega=1$, any finite value of $g$ will lead to mixing between all sectors. Consequently, no single mode will be close to a product state and there will be no revivals. This underlines the importance of non-zero static detuning. The situation is similar when $\Delta_0=(2n+1)\omega/2$, but in this case the $\Omega=0$ Floquet modes split into two groups with same quasienergies (even and odd numbers of excitations). Finite $g$ then causes mixing within these two groups, leading to the appearance of two arcs in the entanglement entropy plots [similar to Fig.~\ref{fig:driven_neel}(c)].
 
In conclusion, the square pulse drive provides analytical understanding of a large part of phase diagram corresponding to high driving amplitudes. However, we note that the agreement between the square pulse and cosine drive is not as good in the high amplitude regime for the polarized state with parameters $\Delta_0=0.64$, $\Delta_m=7.55$, $\omega=2.90$, recall Fig.~\ref{fig:square_pulse}(d).  Nevertheless, the square pulse still accurately captures the revivals and the comoving fidelity is higher than for the delta function pulse protocol. 

\subsection{High-frequency regime}

Although the optimal set of parameters for the N\'eel state is deeply in the non-perturbative regime with $\Omega=1$, $\Delta_0=1.15$, $\Delta_m=2.67$ and $\omega=2.72$ of the same order of magnitude, in Fig.~\ref{fig:frequency_neel_pol} we have seen that it is also possible to somewhat stabilize the revivals by driving with relatively high frequencies. If we follow the same peak by reoptimizing the static detuning and amplitude while increasing the frequency, as in Fig.~\ref{fig:frequency_neel_pol}(a), the resulting driving parameters still result in a subharmonic response and its Floquet modes are split into two arcs. This suggests that we can obtain useful insights into the revival stabilization mechanism by studying the dynamics using the high-frequency expansion. 

\begin{figure}[tbh]
 \includegraphics[width=0.49\textwidth]{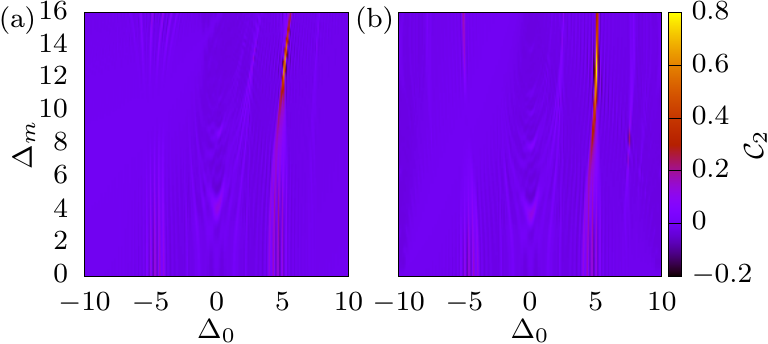}
 \caption{Quality of the subharmonic response, $\mathcal{C}_2$ in Eq.~\eqref{eq:newcost}, for the N\'eel initial state and $\omega=10$. (a) Third order Floquet Hamiltonian in the high frequency regime $H^{(3)}_{F,\omega}$. (b) Time-dependent Hamiltonian $H(t)$ with the square pulse driving scheme.
 }\label{fig:floq_ham_high_freq}
 \end{figure} 

In Appendix~\ref{a:high_frequency} we derive the following third order Floquet Hamiltonian for the square pulse driving protocol in the limit of high driving frequency:
\begin{eqnarray}\label{eq:hF}
 H^{(3)}_{F,\omega}&=&\Omega\left[1+\frac{\Delta_mT^2(3\Delta_0-\Delta_m)}{96}\right]\sum_iP_{i-1}X_iP_{i+1}\nonumber\\
 &-&\Delta_0\sum_in_i+\frac{\Omega^2\Delta_mT^2}{16}\sum_iP_{i-1}Z_iP_{i+1}\\
 &+&\frac{\Omega^2\Delta_mT^2}{16}\sum_iP_{i-1}\big(\sigma^+_{i}\sigma^-_{i+1}+\sigma^-_{i}\sigma^+_{i+1}\big)P_{i+2}.\nonumber 
\end{eqnarray}
In addition to the renormalized PXP term and time averaged detuning $\Delta_0$, this Hamiltonian also contains a diagonal PZP term and constrained nearest-neighbor hopping terms.
Note that the Floquet Hamiltonian only describes stroboscopic evolution at times $t=nT$, where $n$ is an integer.
We have numerically confirmed that the first three orders are enough to correctly reproduce the dynamics of the first few revivals at $\omega=10$. Further increase in frequency makes the agreement even better.
 
 In Fig.~\ref{fig:floq_ham_high_freq} we plot the subharmonic cost function,  Eq.~\eqref{eq:newcost}, for the N\'eel state at $\omega = 10$, and compare the evolution using $H^{(3)}_{F,\omega}$ and the actual square pulse driving protocol. The high-frequency Floquet Hamiltonian correctly reproduces most features of this phase diagram and approximately locates the narrow Type-1 optimal parameter regime (around $\Delta_0\approx5$, $\Delta_m\approx13$).
 At the optimal parameters, the dominant term in $H^{(3)}_{F,\omega}$ is the static detuning, while PZP and hopping are relatively small. The energy spectrum thus resembles that of a strongly detuned PXP model, with several disconnected energy bands defined by their numbers of excitations and the gaps between them approximately equal to $\Delta_0$. If we want to obtain a subharmonic response, this gap should be half the driving frequency, so we expect the optimal detuning to be around $\Delta_0=\omega/2$, which is indeed the case here. The best revivals will be obtained when the energy bands are very narrow, which will make the eigenstates that have the highest overlap with the N\'eel state almost exactly equidistant. The widths of individual bands are controlled by the other three terms. Assuming that $PXP$, $PZP$ and $P\sigma^{\pm}\sigma^{\mp}P$ are approximately of the same order of magnitude and minimizing the sum of their coefficients, $T^2(3\Delta_0\Delta_m-\Delta_m^2+12\Delta_m)/96$, at $\Omega=1$ and fixed $T$ and $\Delta_0$, results in the optimal driving amplitude 
 \begin{eqnarray}\label{eq:deltamopt}
  \Delta_m=3\Delta_0/2+6,
 \end{eqnarray}
which gives $\approx 13.5$ -- close to the numerically obtained optimal value.
 
 We can reproduce Fig.~\ref{fig:frequency_neel_pol}(a) for the square pulse drive (not shown) and compare the dependence of optimal $\Delta_0$ and $\Delta_m$ on the driving frequency with our analytical predictions from the previous paragraph. The optimal static detuning indeed closely follows the relation $\Delta_0(\omega)=\omega/2$, even at relatively low frequencies ($\omega\gtrapprox5$). As expected since the high frequency expansion is no longer valid in that regime, there are some deviations from this rule at lower frequencies, for example $\Delta_0\approx 0.42\omega$ at $\omega=2.72$. Regarding the relation for the optimal driving amplitude $\Delta_m(\omega)=3\omega/4+6$ in Eq.~(\ref{eq:deltamopt}), the numerical results start to approach this line at higher frequencies, $\omega>10$. Since we have already established that the optimal values of static detuning are approximately equal for the cosine drive and square pulse, while the driving amplitude is around 20\% lower for the square pulse, we can use these analytical results to estimate the Type-1 optimal driving parameters for the cosine driving protocol.

\section{Relation to quantum many-body scars}\label{s:scars}

 \begin{figure*}[tbh]
 \includegraphics[width=0.95\textwidth]{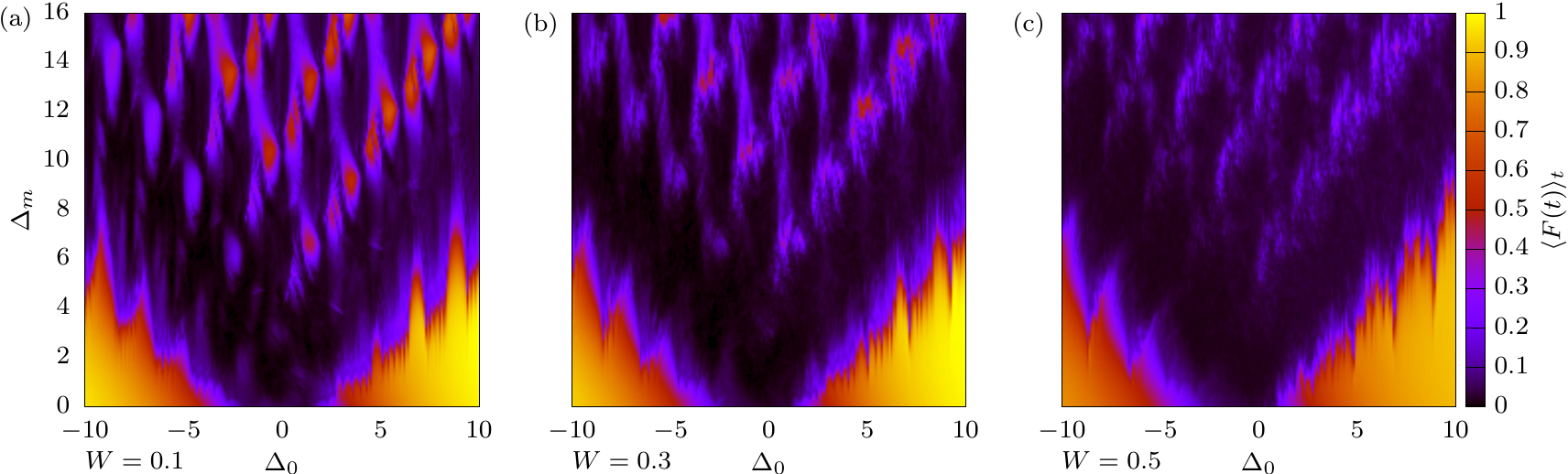}
 \caption{PXP model perturbed by PXPXP perturbation 
 from Eq.~\eqref{eq:pert_nn}. The color scale shows the average fidelity $\langle F(t)\rangle_t$ in a time window [0,100] for the initial N\'eel state. The driving frequency is fixed to $\omega=2$. (a) $W=0.1$. (b) $W=0.3$. (c) $W=0.5$.
 }\label{fig:pert_nn}
 \end{figure*}

Finally, we come to the important set of questions concerning the relation between QMBS and the enhancement of fidelity revivals by periodic driving. We focus on two questions in particular: (i) Is it still possible to stabilize the revivals from some initial state if the PXP part of the Hamiltonian in Eq.~\eqref{eq:ht} is replaced by some other, fully ergodic Hamiltonian?  (ii) How do we prove that the optimally-driven system is  ``scarred"? 
The second question, in particular, is non-trivial as many probes of QMBS in static systems do not directly translate to the driven case. For example, QMBS are typically diagnosed as a band of atypical eigenstates at equidistant energies. Since the driven model is time dependent, the Floquet modes assume the role of the eigenstates. However, in all the cases studied here, there is no band of special Floquet modes in the quasienergy spectrum of the Floquet Hamiltonian. Instead, there is only one or in some cases two special modes, which have unusually low entanglement entropy and very high overlap with the initial state. Also, it could be said that these modes are typically located at the edges of the quasienergy spectrum rather than the middle of it,  see Fig.~\ref{fig:driven_neel}(c). However, one must keep in mind it is hard to rigorously define the ``edge'' when the quasienergy spectrum is periodic. In light of this difference between driven and static systems, it is pertinent to ask if the dynamics in the driven case is still ``scarred" in the sense that it represents merely an enhancement of the effect already present in the static case. In this section we address these questions.

\subsection{The effect of perturbation}

QMBS can be destroyed by applying suitable perturbations to the PXP model~\cite{Turner2018b}. The destruction of QMBS was found to coincide with the overall level statistics of the model converging faster to the Wigner-Dyson distribution, signaling faster thermalization. Here we study the PXPXP perturbation and apply the usual spatially homogeneous cosine driving protocol, Eq.~\eqref{eq:cos_drive}, to the perturbed PXP Hamiltonian,
\begin{eqnarray}\label{eq:pert_nn}
H'_\mathrm{PXP}(W) &=& H_\mathrm{PXP} + W\sum_i P_{i-1}X_{i}P_{i+1}X_{i+2}P_{i+3},\ \\
H'(t) &=& H'_\mathrm{PXP}(W) - \Delta(t)\sum_i n_{i}.\label{eq:ht_pert}
\end{eqnarray}
The PXPXP perturbation is off-diagonal, thus it changes the connectivity of the Hilbert space.

In Fig.~\ref{fig:pert_nn} we scan the $(\Delta_0,\Delta_m)$ parameter space for three different perturbation strengths, while the driving frequency is fixed at $\omega=2$. Similar to Fig.~\ref{fig:fidelity_scan_neel}, the color corresponds to the average fidelity $\langle F(t)\rangle_t$ between $t=0$ to $t=100$, such that the bright regions correspond to parameter regimes where we expect robust revivals. The perturbation destroys some of the optimal parameter regimes, mostly those close to the diagonals $\Delta_0=\pm\Delta_m$. For example, the main Type-1 optimal regime for the N\'eel state belongs to this region, as shown in Fig.~\ref{fig:fidelity_scan_neel}(a) for $\omega=2.72$. The higher-amplitude optimal parameter regions (Type-2) are affected by the perturbation and their position, shape and size all depend on the perturbation strength. However, these bright regions are still present even when the perturbation is fairly strong ($W=0.5$).
Those regions are related to Hilbert space fragmentation, see Fig.~\ref{fig:driven_polarized}. This allows us to conclude that the Floquet Hamiltonian fragmentation mechanism is not specific to the PXP model, which is in line with the discussion from Sec.~\ref{ss:arcs}. However, a possibility remains that the original, weakly nonergodic PXP model is in some way important for the revival enhancement in the low-amplitude driving regime for the N\'eel state (Fig.~\ref{fig:driven_neel}).

Similar behavior was also observed with different types of perturbations, such as next-nearest neighbor interaction $W_1\sum_i n_{i}n_{i+2}$, or constrained nearest neighbor hopping $W_2\sum_i P_{i-1}\left(\sigma^{+}_{i}\sigma^{-}_{i+1}+\sigma^{-}_{i}\sigma^{+}_{i+1}\right)P_{i+2}$. It is, however, interesting to note that, while destroying scars for the $\mathbb{Z}_2$ state, the next-nearest neighbor interaction can actually create scars for the polarized state, which is reminiscent of the effects of static detuning~\cite{Su2022}. The stabilization of revivals from the polarized state using periodic driving works even better (quantified by higher $\mathcal{C}_1$) in a certain range of nonzero values of $W_1$ that coincides with the existence of quantum many-body scars for this state in the static system.

\subsection{Trajectory in the Hilbert space}\label{ss:trajectory}

\begin{figure}[bth]
 \includegraphics[width=0.49\textwidth]{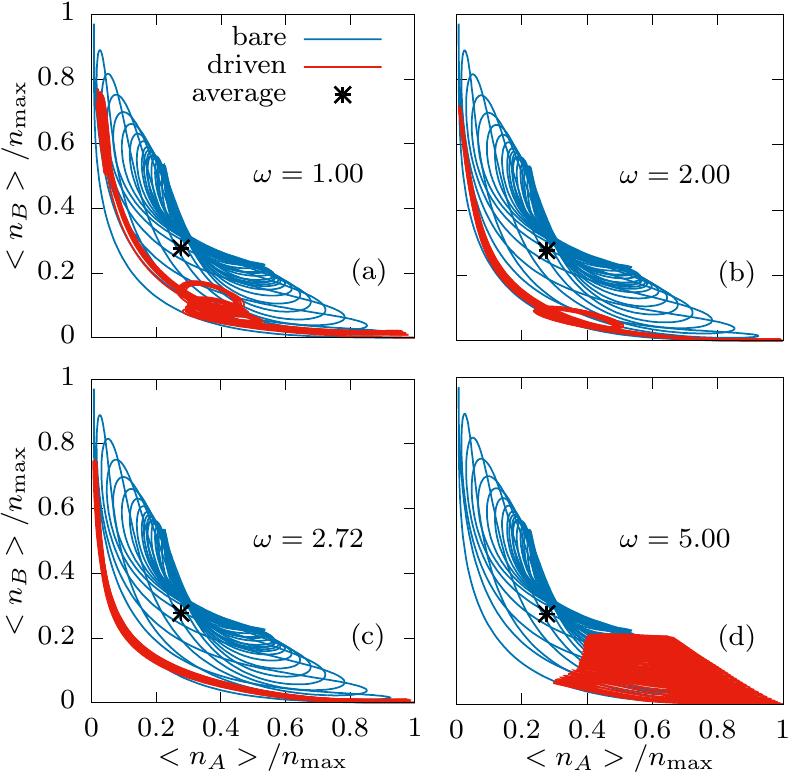} 
 \caption{Trajectory in the Hilbert space. N\'eel state without (blue) and with optimal driving (red). $N=24$. The maximal number of excitations per sublattice is $n_\mathrm{max}=N/2$. 
 (a) $\omega=1.00$, $\Delta_0=-0.65$, $\Delta_m=8.31$. (b) $\omega=2.00$, $\Delta_0=-1.17$, $\Delta_m=4.81$. (c) $\omega=2.72$, $\Delta_0=1.15$, $\Delta_m=2.67$. (d) $\omega=5.00$, $\Delta_0=2.34$, $\Delta_m=6.85$.
 }\label{fig:trajectory_neel}
 \end{figure}

Another way to relate the reviving dynamics in the driven system with QMBS in the static case is to examine the Hilbert space trajectory of the N\'eel state, with and without driving. We can visualize this trajectory by plotting the expectation values of the (normalized) numbers of excitations on the even and odd sublattice, $\langle n_A\rangle$ and $\langle n_B\rangle$ respectively, see Fig.~\ref{fig:trajectory_neel}. The two N\'eel states only have excitations on one of the sublattices, so they correspond to the bottom right (1,0) and top left (0,1) corner, while the polarized state with no excitations is in the bottom left corner (0,0). All other states on the main diagonal or above it are forbidden due to the PXP blockade constraint.

Up to this point, for the N\'eel state we have mostly used the driving frequency $\omega=2.72$, which is approximately twice the frequency of revivals in the static PXP model ($\omega\approx2.66$). Similarly, the optimal driving frequency for the polarized state in the low-amplitude regime is close to the frequency of revivals in the detuned PXP model.
However, we were also able to stabilize the revivals at driving frequencies that are not particularly close to the original revival frequency. Upon first glance, this fact does not seem to support the claim that the driven system hosts QMBS, however we will show below that driving with frequencies close to $\omega=2.72$ leads to Hilbert space trajectories with some special properties.

It is possible to obtain high and long-lived revivals from the N\'eel state at frequencies significantly below $\omega=2.72$. Two examples of such trajectories are shown by the red lines in Figs.~\ref{fig:trajectory_neel}(a) and \ref{fig:trajectory_neel}(b), for driving frequencies $\omega=1$ and $\omega=2$ respectively, while the trajectory in the static model is plotted in blue.
For comparison, the trajectory for $\omega=2.72$ is shown in Fig.~\ref{fig:trajectory_neel}(c).
In this special case, the trajectory precisely repeats the approximate first revival period of the undriven case. It could therefore be said that driving with optimal parameters stabilizes the scarred dynamics of the pure PXP model. 
This suggests that the true optimal driving frequency is indeed related to quantum many-body scars in the static model.
The most significant difference between this trajectory and the previous two is that the low-frequency trajectories contain loops. The number of loops grows as the driving frequency decreases. Although the trajectory can still be made to oscillate between the two N\'eel states at lower frequencies, the intermediate dynamics differs from that in the static PXP model.

\begin{figure*}[tbh]
 \includegraphics[width=0.31\textwidth]{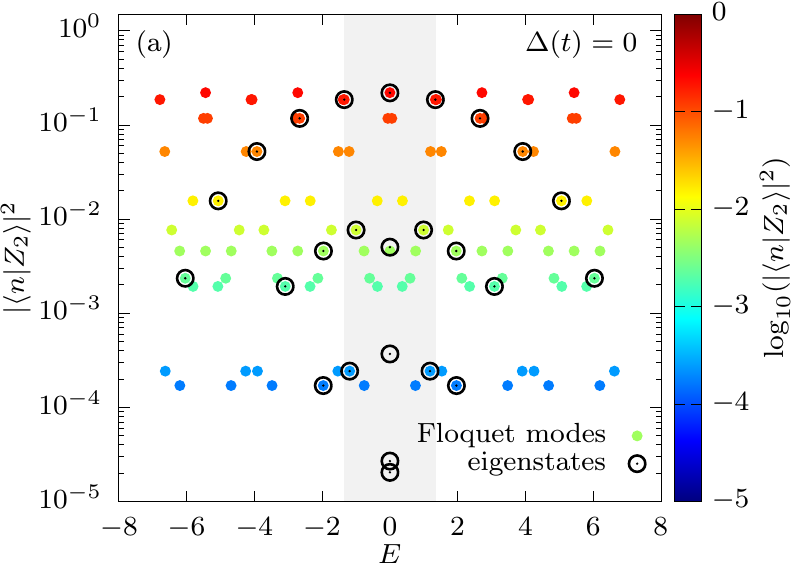}
 \includegraphics[width=0.31\textwidth]{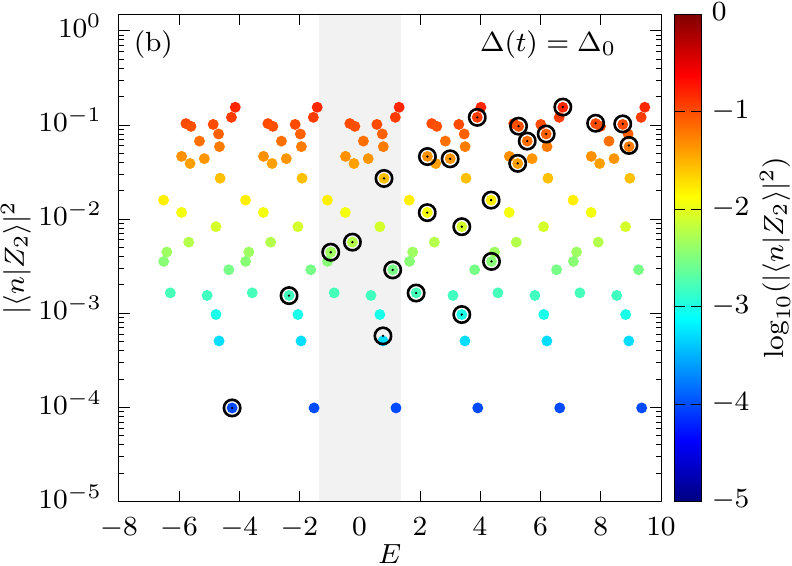}
 \includegraphics[width=0.31\textwidth]{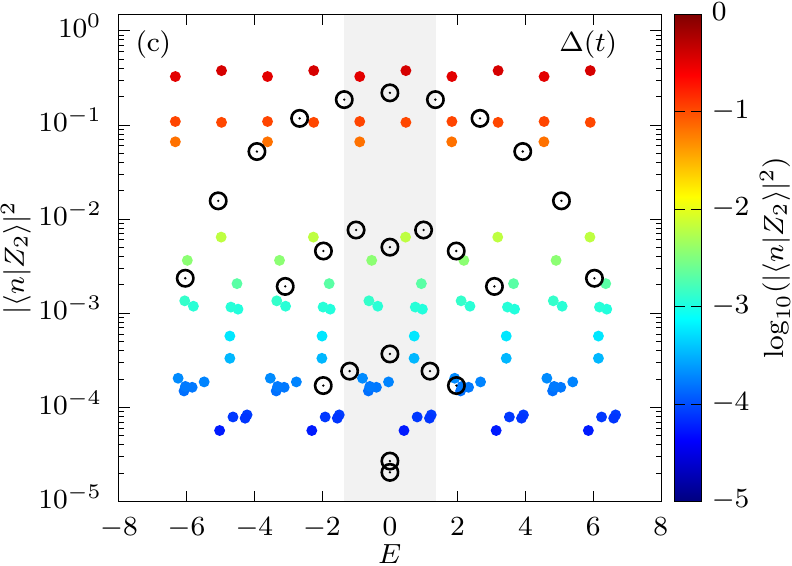}\\
 \includegraphics[width=0.47\textwidth]{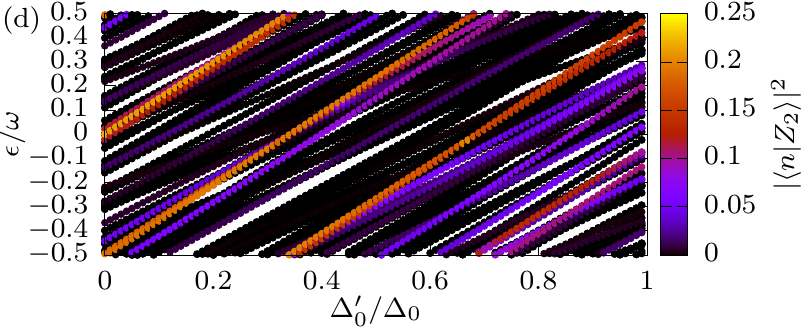} 
 \includegraphics[width=0.47\textwidth]{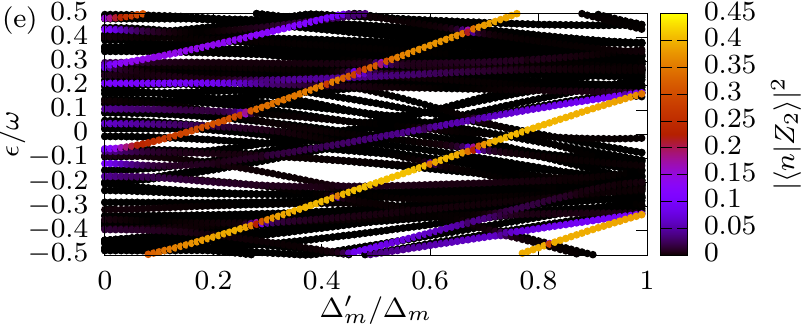}\\
 \includegraphics[width=0.47\textwidth]{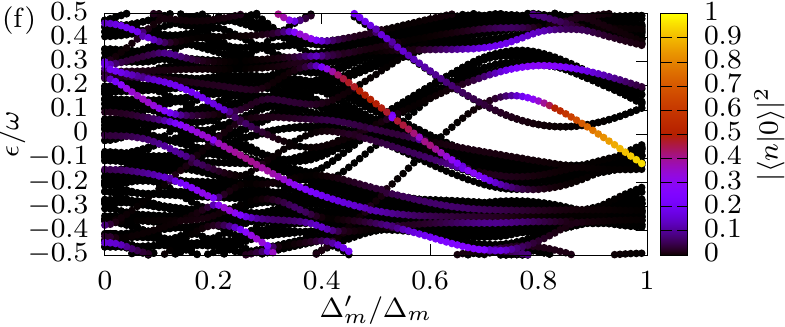}
 \includegraphics[width=0.47\textwidth]{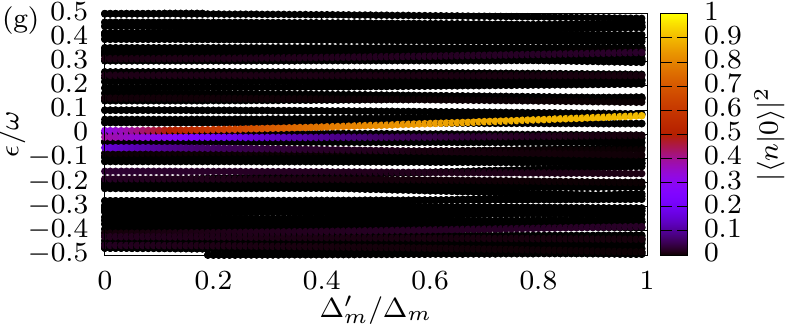}
 \caption{
 Overlap of the Floquet modes with the initial state, $N=10$. (a) Bare PXP model (N\'eel, $\omega=2.72$). Floquet modes (color) versus eigenstates (black circles). The shaded area marks the $[-\omega/2,\omega/2]$ interval of the Floquet spectrum. (b) Same, but with static detuning $\Delta_0=1.15$. (c) Same, but optimally driven ($\Delta_0=1.15$, $\Delta_m=2.67$). The eigenstates of the bare PXP model are shown for comparison. (d) Interpolation between bare and detuned (N\'eel). (e) Interpolation between detuned and optimally driven (N\'eel). 
 (f) Interpolation between detuned and optimally driven (polarized, high amplitude, $\Delta_0=0.64$, $\Delta_m=7.55$, $\omega=2.90$).
 (g) Interpolation between detuned and optimally driven (polarized, low amplitude, $\Delta_0=1.68$, $\Delta_m=-0.50$, $\omega=3.71$).
 }\label{fig:interpolation}
 \end{figure*} 

However, the revivals cannot be stabilized for any arbitrarily chosen driving frequency. When the frequency is too high, for example $\omega=5$ in Fig.~\ref{fig:trajectory_neel}(d), the revivals decay and the trajectory explores only one corner of the Hilbert space, staying in the vicinity of the initial state. This could be better understood using the square pulse approximation of the driving protocol from Sec.~\ref{s:square}. Close to integer multiples of the driving period, the drive is in the high-detuning regime ($\Delta_0+\Delta_m$)
where the spectrum is split into bands and the N\'eel state is approximately one of the eigenstates. Frequent returning to this regime therefore prevents the wave function from moving away from the N\'eel state. The intermediate low-detuning regime ($\Delta_0-\Delta_m$) needs to be long enough in order for the trajectory to reach the other side of the Hilbert space, and its detuning low enough so that the system is closer to the pure PXP model. In other words, the frequency needs to be low enough for the driven system to oscillate between the two N\'eel states.

Two remarks are in order. First, driving with high-amplitude optimal parameters for the polarized state can also produce high revivals from the N\'eel state. This is not surprising since in this case there are two Floquet modes with high overlap with the N\'eel states [see Fig.~\ref{fig:driven_polarized}(f)], but the wave function is in that case ``stuck'' in one corner of the Hilbert space. The trajectory is similar to Fig.~\ref{fig:trajectory_neel}(d), albeit more narrow.
Second, representing the Hilbert space trajectory in the $\langle n_A\rangle$-$\langle n_B\rangle$ plane is not suitable for the polarized state, since both that state and the full (uniformly driven) Hamiltonian are symmetric with respect to the two sublattices, thus $\langle n_A\rangle=\langle n_B\rangle$ at all times, i.e., the trajectory starting from the polarized state always stays on the diagonal. In Appendix~\ref{a:trajectory_polarized} we discuss an alternative way to plot the trajectory for the polarized state.

\subsection{From scarred eigenstates to Floquet modes}

Finally, we would like to explore the connection between scarred eigenstates in the static PXP model and the two $\pi$-paired Floquet modes that have high overlap with the N\'eel state. To this end, we interpolate between the static [$\Delta(t)=0$]
and optimally driven system [$\Delta(t)=\Delta_0+\Delta_m\cos(\omega t)$] 
by first increasing the static detuning from 0 to $\Delta_0$ while keeping $\Delta_m=0$, and then increasing the driving amplitude to its optimal value. The driving frequency is kept constant throughout this process ($\omega=2.72$). The Floquet modes can be formally computed even in the $\Delta_m\rightarrow0$ limit. In this case the Floquet modes should be exactly equal to the eigenstates of the static system. However, the Floquet quasienergy spectrum is periodic with periodicity $\omega$, and we therefore need to ``unfold'' it in order to match the Floquet modes with their corresponding eigenstates. This situation is shown in Fig.~\ref{fig:interpolation}(a) where we plot the overlap of the PXP eigenstates and Floquet modes with the N\'eel state. For clarity, we use a small system size $N=10$. The shaded region marks the original $[-\omega/2,\omega/2]$ interval of the Floquet spectrum, which is translated by $\omega$ several times so that the full energy spectrum is covered.

Next, the Floquet modes are computed for various values of static detuning $\Delta'_0$ and their quasienergies and overlap with the N\'eel state are plotted in Fig.~\ref{fig:interpolation}(d). The highest overlap states of the bare PXP model are continuously deformed into the highest-overlap states of the detuned model at $\Delta_0=1.15$, which are also shown in Fig.~\ref{fig:interpolation}(b). Similarly, these two states are then transformed into the two special Floquet modes by increasing the driving amplitude $\Delta'_m$ from $0$ to $\Delta_m=2.67$, see Fig.~\ref{fig:interpolation}(e). To summarize, it is possible to establish a connection between the scarred eigenstates of the static PXP model [for example, $\epsilon=0$ and $\epsilon\approx \omega/2$ at $\Delta'_0/\Delta_0=0$ in Fig.~\ref{fig:interpolation}(d)] and the $\pi$-paired modes that are responsible for near-perfect revivals in the optimally driven model [two highest overlap modes at $\Delta'_m/\Delta_m=1$ in Fig.~\ref{fig:interpolation}(e)]. Driving leads to scarred Floquet modes, which are more separated from the bulk than in the original bare PXP model, see Fig.~\ref{fig:interpolation}(c). The $\pi$-pairing 
was already present in the static model and was never broken during the interpolation procedure.

The interpolation between the detuned and optimally driven system was repeated for the polarized state and the results are shown in Figs.~\ref{fig:interpolation}(f) and \ref{fig:interpolation}(g). Unlike for the N\'eel state, it is not possible to continuously deform the eigenstate of the static detuned model into the the high overlap Floquet mode of the driven system in the high-amplitude regime, see Fig.~\ref{fig:interpolation}(f). However, as shown in Fig.~\ref{fig:interpolation}(g), the low-amplitude driving regime is different. In that case, one of the scarred eigenstates indeed transforms into the highest overlap Floquet mode.

\section{Conclusions}\label{s:conclusions}

In this work we studied the periodically driven PXP model in order to gain a better understanding of how driving can enhance the revivals that are present in the static model due to QMBS. 
We introduced a square pulse approximation for the cosine driving protocol that was used in earlier experiments, and demonstrated that this approximation much more accurately reproduces the dynamics of the cosine drive than the previously studied delta function pulse drive, while still being  amenable to analytical treatment. This allowed us to analytically predict the optimal driving parameters in the high-amplitude and high-frequency regimes. We were also able to modify the cosine driving protocol in such a way that any initial product state can be made to revive until late times. 

We identified two distinct classes of optimal parameter regimes, which were dubbed Type-1 and Type-2. The second class is related to the splitting of the Floquet Hamiltonian into disconnected sectors, where a single Floquet mode belongs to a one-dimensional sector and has high overlap with one of the product states. The sectors consist only of Floquet modes with a certain number of excitations, thus the drive, being the largest energy scale, produces an emergent U(1) symmetry, similar to prethermalization scenario~\cite{Abanin2017}. Type-2 mechanism of revival creation and stabilization is well understood, relatively trivial and likely not connected with QMBS. The optimal driving parameter regimes were found for completely arbitrary initial product states, and were shown to be insensitive to perturbations that destroy quantum many-body scars in the original PXP model.

The other class of optimal driving parameters (Type-1) is more interesting, but also more challenging to study analytically due to it being in the non-perturbative regime. This revival enhancement mechanism was so far only observed for the N\'eel state and is destroyed by perturbations that also destroy QMBS. Visualising the trajectory of the evolved state in the Hilbert space and comparing the static and driven cases reveals that optimal driving stabilizes this trajectory -- the system is made to precisely repeat the dynamics of the first revival period of the undriven model. When the system is initialized in one of the two N\'eel states, the response to driving is subharmonic, meaning that the revival frequency is half the driving frequency~\cite{Maskara2021}. 
However, the subharmonic response becomes harmonic when the initial state is a translation-invariant superposition of the two N\'eel states. The origin of the subharmonic response is related to the presence of two $\pi$-paired Floquet modes in the quasienergy spectrum of the optimally driven system. These two atypical modes have very high overlap with the N\'eel states and can be thought of as equivalents of scarred eigenstates in a periodically driven system.

An interesting direction for future work is the extension of driving protocols to other models that host inexact QMBS states and revival dynamics. This would involve finding a suitable driving term whose ground state is the scarred initial state. While Type-2 parameters (high-amplitude regime) are expected to trivially work even in the absence of scars, a more interesting question is whether Type-1 phenomenology exists beyond the PXP model.

 \begin{figure*}[htb]
 \includegraphics[width=0.7\textwidth]{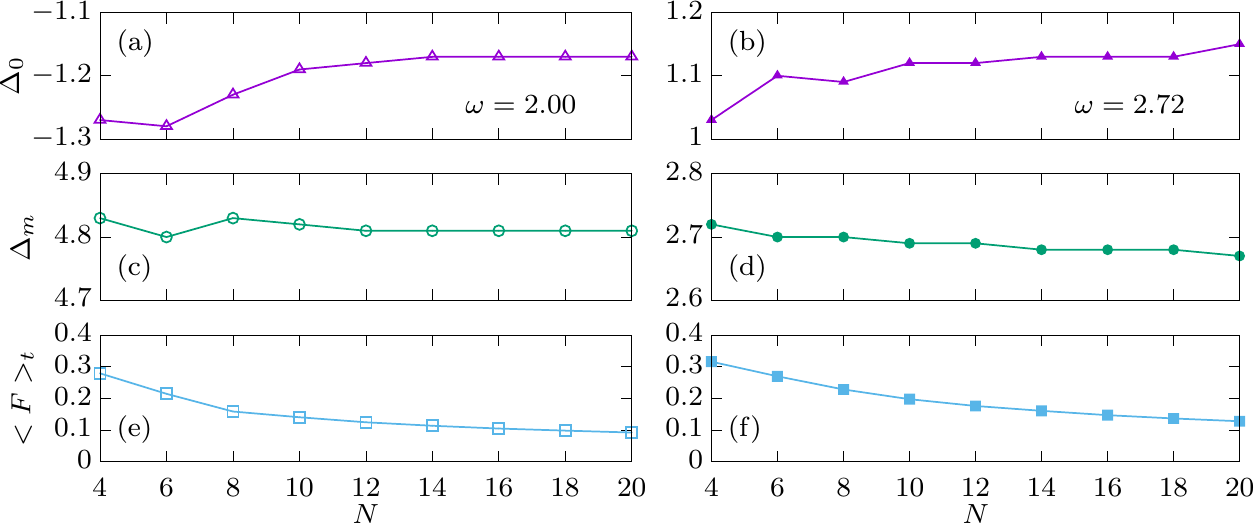}
 \caption{System size scaling of the optimal driving parameters and the average fidelity for the N\'eel state. Left: $\omega=2.00$ (one-loop trajectory). Right: $\omega=2.72$ (ideal trajectory). (a)-(b) Static detuning $\Delta_0$. (c)-(d) Driving amplitude $\Delta_m$. (e)-(f) Average fidelity from $t=0$ to $t=100$, $\langle F\rangle_t$. }\label{fig:size_scaling}
 \end{figure*} 

\acknowledgments

We would like to thank Hui Sun, Zhao-Yu Zhou, Bing Yang, Zhen-Sheng Yuan and  Jian-Wei Pan for collaboration on a related project.  A.H., J.-Y.D., and Z.P.~acknowledge support by EPSRC grant EP/R513258/1 and by the Leverhulme Trust Research Leadership Award RL-2019-015. 
A.H.~acknowledges funding provided by the Institute of Physics Belgrade, through the grant by the Ministry of Education, Science, and Technological Development of the Republic of Serbia. Part of the numerical simulations were performed at the Scientific Computing Laboratory, National Center of Excellence for the Study of Complex Systems, Institute of Physics Belgrade. J.C.H.~acknowledges funding from the European Research Council (ERC) under the European Union's Horizon 2020 research and innovation program (Grant Agreement no 948141) — ERC Starting Grant SimUcQuam, and by the Deutsche Forschungsgemeinschaft (DFG, German Research Foundation) under Germany's Excellence Strategy -- EXC-2111 -- 390814868.

\appendix
\section{System size scaling}\label{a:scaling}

In this Appendix we examine the system size scaling of several important quantities. We fix the driving frequency and scan the $(\Delta_0,\Delta_m)$ parameter space in the vicinity of two Type-1 peaks, $(\Delta_0,\Delta_m,\omega)=(-1.17,4.81,2.00)$ and $(\Delta_0,\Delta_m,\omega)=(1.15,2.67,2.72)$, for different system sizes $N$. The results are presented in Figs.~\ref{fig:size_scaling}(a)-\ref{fig:size_scaling}(d), showing that the optimal driving parameters are relatively stable for $N\geq10$. Additionally, we plot the maximal average fidelity $\langle F\rangle_t$ as a function of system size in Figs.~\ref{fig:size_scaling}(e) and \ref{fig:size_scaling}(f). Based on these results, we predict that the same parameters will still be optimal and produce revivals of finite height in relatively large systems. Such system sizes are out of reach of numerical simulations based on exact diagonalization, but could be realized experimentally. For example, a recent experiment has realized the PXP model with approximately 50 sites using a Bose-Hubbard quantum simulator \cite{Su2022}.

  \begin{figure}[htb]
 \includegraphics[width=0.35\textwidth]{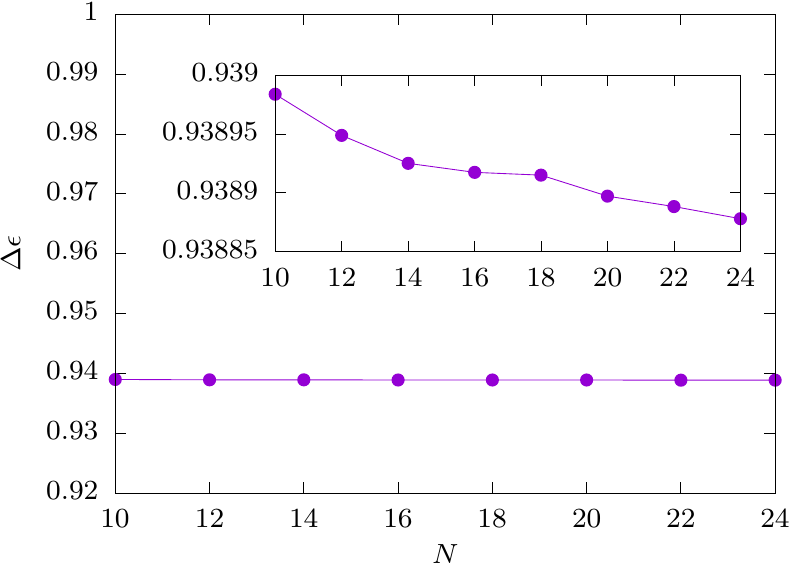}
 \caption{System size scaling of the quasienergy gap between the two lowest entropy Floquet modes. Optimal parameters for the polarized state ($\Delta_0=0.64$, $\Delta_m=7.55$, $\omega=2.90$). The inset shows the same data with a different $y$-axis range.
 }\label{fig:size_scaling_gap_pol}
 \end{figure} 
 
 In Fig.~\ref{fig:driven_polarized}(f) we showed the quasienergy spectrum of the Floquet modes that correspond to the optimally driven polarized state in the high-amplitude regime. This spectrum is split into separate bands according to the number of excitations, with a single Floquet mode approximately equal to the polarized state belonging to a one-dimensional sector (0 excitations) and separated from the neighboring sector (1 excitation) by a gap $\Delta\epsilon$. In Fig.~\ref{fig:size_scaling_gap_pol} we  show that this gap is almost independent of system size. The inset zooms in on the data points in order to show that weak exponential decay is nevertheless present. If this trend continues, we estimate a finite gap even at sizes  $N\approx300$, implying that it would still be possible to create fidelity revivals in such large systems by applying the same driving parameters found in smaller systems.
 
 \begin{figure}[tbh]
 \includegraphics[width=0.4\textwidth]{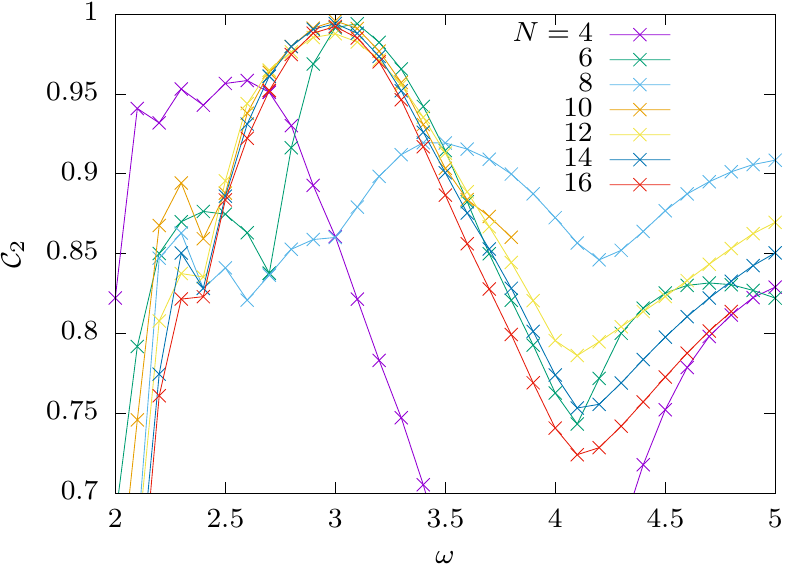}
 \caption{Cost function $\mathcal{C}_2$ in Eq.~(\ref{eq:newcost}) with $M=20$ as a function of frequency for different system sizes. ``Main'' Type-1 peak, which passes through the point $(\Delta_0, \Delta_m, \omega)=(1.15, 2.67, 2.72)$.
 }\label{fig:frequency_neel_size}
 \end{figure}
 
 We also investigated the dependence of the optimal frequency on system size. We use the same procedure as in Fig.~\ref{fig:frequency_neel_pol} and plot the cost function $\mathcal{C}_2$ as a function of driving frequency in Fig.~\ref{fig:frequency_neel_size}. The optimal frequency window is quite stable for $N\geq10$, although it becomes slightly narrower as the system size increases.

\section{Spatially inhomogeneous drive}\label{a:inhomogeneous}

\begin{figure}[htb]
 \includegraphics[width=0.25\textwidth]{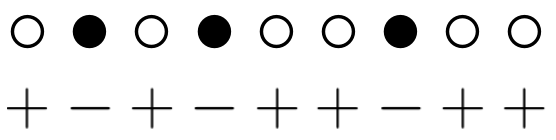}
 \caption{Modified driving protocol for an arbitrary product state. The protocol is now site-dependent, i.e., each site is driven by a term $(-1)^{n_{0i}} \Delta(t)$ whose sign depends on whether or not the site initially hosts an excitation.  
 }\label{fig:pm_drive}
 \end{figure}

Based on previous results, we now introduce a generalized driving protocol, which can stabilize revivals from \emph{any} product state. A key observation is that the two initial states whose revivals can be stabilized by driving (N\'eel and polarized) are the ground and the highest-excited state of the spatially homogeneous driving term in Eq.~\eqref{eq:ht}, since these two states have the highest and lowest possible number of excitations allowed by the blockade. In order to achieve a similar scenario for other states, the driving term should be chosen as spatially-dependent,
\begin{eqnarray}\label{eq:inhomogeneous}
H(t) = H_\mathrm{PXP} + \Delta(t)\sum_i (-1)^{n_{0i}}n_{i},
\end{eqnarray}
where $n_{0i}=\langle\psi(0)\lvert n_i\rvert\psi(0)\rangle$ is determined by the initial state. In this protocol, each site is driven by $\pm \Delta(t)$, where the sign depends on whether the site initially contains an excitation or not, see Fig.~\ref{fig:pm_drive} for an illustration. In this way, the initial state becomes the (non-degenerate) ground state of the driving term.
 
Using the inhomogeneous driving protocol, we were able to generate very high and long lived revivals for any initial state, even randomly chosen ones with no symmetries. The driving parameters $\Delta_0$, $\Delta_m$ and $\omega$ need to be separately optimized for each initial state. To this end, we use the simulated annealing optimization algorithm, which limits the system sizes we can reach. However, when the initial state is sufficiently symmetric, for example $\mathbb{Z}_2$, $\mathbb{Z}_3$, $\mathbb{Z}_4$ and combinations of such states, it is sufficient to find the optimized parameters in a smaller system and the same driving parameters can be applied to the same state in larger systems, as was the case for $\mathbb{Z}_2$ with uniform driving. 

\begin{table}
\begin{tabular}{ |c|c|c|c|c| }
\hline
Initial state & $\Delta_0$ & $\Delta_m$ & $\omega$ & Floquet spectrum\\
\hline
\multirow{2}{4em}{$\mathbb{Z}_2$} & -1.61 & 1.83 & 2.83 & three arcs\\ 
& 0.45 & 5.89 & 2.16 & multiple arcs\\ 
\hline
\multirow{2}{4em}{$\mathbb{Z}_3$} & -0.26 & 4.14 & 2.24 & one arc \\ 
& 0.70 & 5.43 & 1.82 & multiple arcs\\ 
\hline
\multirow{1}{4em}{$\mathbb{Z}_4$} & 0.48 & 5.64 & 2.21 & multiple arcs \\
\hline
\multirow{1}{4em}{$\mathbb{Z}_2\mathbb{Z}_3$} & 0.71 & 5.73 & 1.90 & multiple arcs \\
\hline
\multirow{1}{4em}{random} & 0.60 & 6.36 & 2.33 & multiple arcs \\
\hline
\end{tabular}
\caption{Optimal driving parameters. 
$\mathbb{Z}_2$ (N\'eel), $\mathbb{Z}_3$ and $\mathbb{Z}_4$ are the states with an excitation on the every second, third and fourth site, respectively.
$\mathbb{Z}_2\mathbb{Z}_3$ is $\ket{{\bullet}{\circ}{\bullet}{\circ}{\circ}{\bullet}{\circ}{\bullet}{\circ}{\circ}\ldots{\bullet}{\circ}{\bullet}{\circ}{\circ}}$ (alternating $\mathbb{Z}_2$ and $\mathbb{Z}_3$), while the randomly chosen state is $\ket{{\bullet}{\circ}{\bullet}{\circ}{\circ}{\circ}{\circ}{\circ}{\bullet}{\circ}{\circ}{\bullet}{\circ}{\circ}{\circ}{\circ}}$. 
}
\label{tab:parameters_inhomogeneous}
\end{table}
  
 \begin{figure}[htb]
 \includegraphics[width=0.45\textwidth]{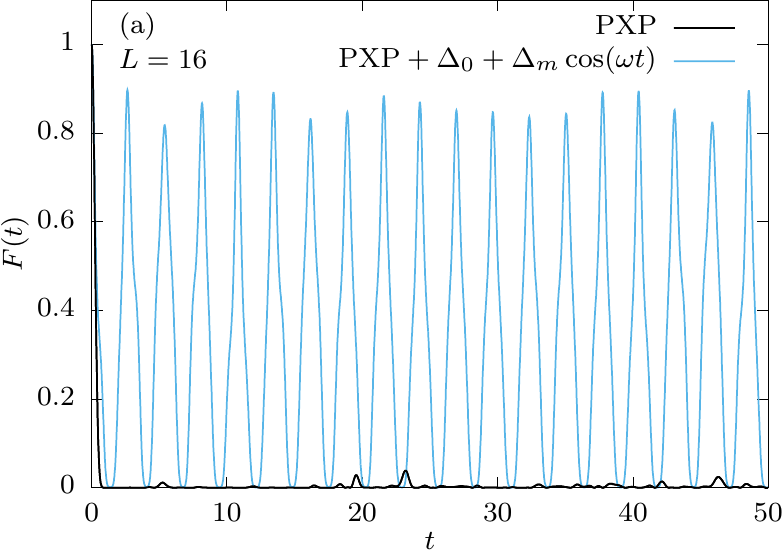}\\
 \includegraphics[width=0.45\textwidth]{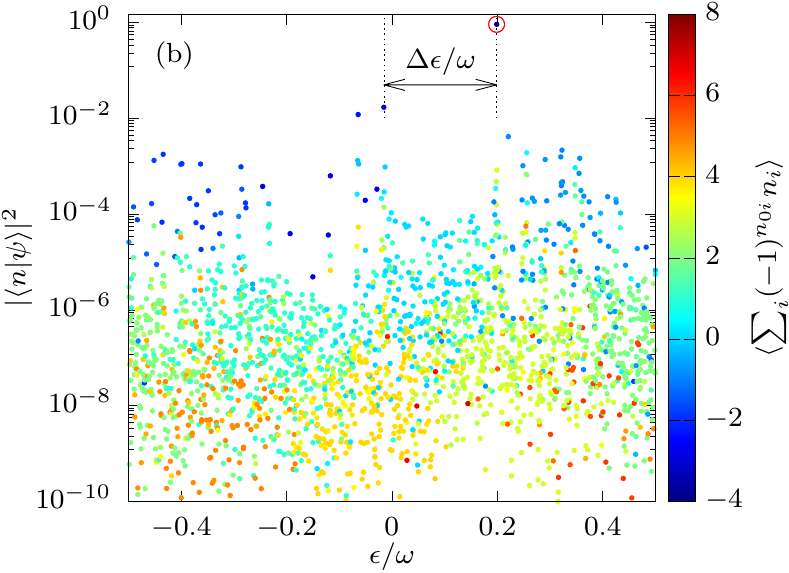}
 \caption{Inhomogeneously driven random state $\ket{{\bullet}{\circ}{\bullet}{\circ}{\circ}{\circ}{\circ}{\circ}{\bullet}{\circ}{\circ}{\bullet}{\circ}{\circ}{\circ}{\circ}}$ ($\Delta_0=0.60$, $\Delta_m=6.36$, $\omega=2.33$). (a) Quantum fidelity. 
 (b) Overlap of the initial state with all the Floquet modes. The color corresponds to $\langle\sum_i (-1)^{n_{0i}}n_{i}\rangle$ for each mode.
 }\label{fig:driven2_random}
 \end{figure}

Some optimal parameter values obtained by simulated annealing are listed in Table~\ref{tab:parameters_inhomogeneous}. It is interesting to note that these values are similar in most cases, with $\lvert\Delta_0\rvert<1$, $\Delta_m\gg1$ and $\omega\approx2$. From these values, we see that the phenomenology is similar to the polarized state in the high-amplitude driving regime [cf.~Figs.~\ref{fig:driven_polarized}(d)-\ref{fig:driven_polarized}(f)]:  the quasienergy spectrum is split into sectors with different values of $\langle\sum_i (-1)^{n_{0i}}n_{i}\rangle$ and there is a single Floquet mode that has very high overlap with the initial state. 
Ultimately, the mechanism of scar enhancement in this case is  very special case of  prethermalization~\cite{Abanin2017}: $\Delta_m$ assumes the role of the large energy scale, which gives rise to an approximate U(1) symmetry. The initial state belongs to a one-dimensional sector and its mixing with other states is suppressed over long times, in accordance with general results in Ref.~\cite{Abanin2017}.

To illustrate the spatially inhomogeneous drive in Eq.~\eqref{eq:inhomogeneous}, we consider a completely random state $\ket{{\bullet}{\circ}{\bullet}{\circ}{\circ}{\circ}{\circ}{\circ}{\bullet}{\circ}{\circ}{\bullet}{\circ}{\circ}{\circ}{\circ}}$. This state has no special properties and quickly thermalizes in the pure PXP model, but driving with optimal parameters results in stable revivals, see Fig.~\ref{fig:driven2_random}(a). It might be hard to see due to a large number of different sectors and relatively small system size, but the quasienergy spectrum is also fragmented in this case. As can be observed in Fig.~\ref{fig:driven2_random}(b), there is again a single high-overlap mode, which belongs to a one-dimensional sector, since it is the only product state that has the minimal possible value of $\langle\sum_i (-1)^{n_{0i}}n_{i}\rangle$. The driving amplitude is the largest energy scale ($\Delta_m=6.36$).

\begin{figure}[htb]
 \includegraphics[width=0.45\textwidth]{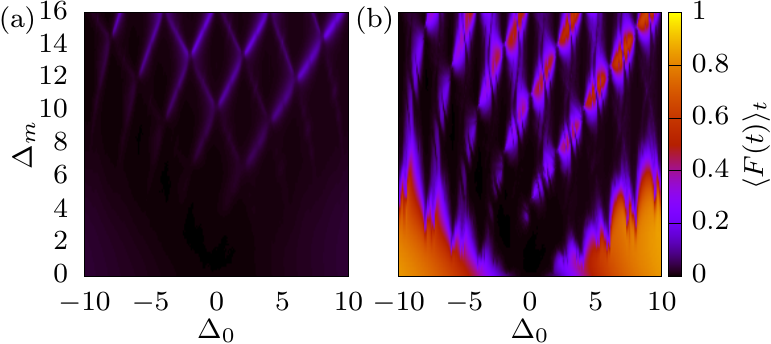}
 \caption{Average fidelity $\langle F(t)\rangle_t$ over time interval $[0, 100]$ for a driven random state $\ket{{\bullet}{\circ}{\bullet}{\circ}{\circ}{\circ}{\circ}{\circ}{\bullet}{\circ}{\circ}{\bullet}{\circ}{\circ}{\circ}{\circ}}$, $\omega=2$. (a) Homogeneous drive, Eq.~\eqref{eq:cos_drive}. 
 (b) Inhomogeneous drive, Eq.~\eqref{eq:inhomogeneous}.
 }\label{fig:scan_random}
 \end{figure}

Finally, in Fig.~\ref{fig:scan_random} we scan the average fidelity cost function over the parameter space for the same randomly chosen state and fixed driving frequency $\omega=2$. For comparison, in Fig.~\ref{fig:scan_random}(a) we use the homogeneous driving protocol from Eq.~\eqref{eq:cos_drive}. In this case, there are no significant peaks in the cost function, unlike for the N\'eel and polarized state in Fig.~\ref{fig:fidelity_scan_neel}. However, as shown in Fig.~\ref{fig:scan_random}(b), the inhomogeneous driving scheme from Eq.~\eqref{eq:inhomogeneous} results in a similar plot as those in Fig.~\ref{fig:fidelity_scan_neel}. As for the polarized state, it contains only Type-2 peaks (no subharmonic response).

\section{High frequency expansion}\label{a:high_frequency}

Here we calculate the effective Floquet Hamiltonian, which governs the stroboscopic dynamics for the three-step square pulse
protocol in Eq.~\eqref{eq:square} under the assumption of high driving frequency ($T\ll 1$). To this end, we perform the expansion in powers of $T$ using the Baker-Campbell-Hausdorff (BCH) series. We have numerically confirmed that higher frequency Type-I peaks can be well captured by a finite order expansion. 
Motivated by this, we analytically calculate the Floquet Hamiltonian up to third order in BCH series. 

We start by writing down the BCH series for the multiplication of three exponential matrices $e^z = e^xe^ye^w$, which gives the expansion of $z$ as follows
\begin{eqnarray}
 z&=&z^{(1)}+z^{(2)}+z^{(3)}+\cdots\nonumber\\
 z^{(1)}&=&x+y+w,\nonumber\\
 z^{(2)}&=&\frac{1}{2}(-wx-wy+xw+xy+yw-yx),\nonumber\\
 z^{(3)}&=&\frac{1}{12}(w^2x+w^2y+wx^2-2wxw-2wxy\nonumber\\
 &&+wy^2-2 wyw+4wyx+x^2w\nonumber\\
 &&+x^2y+xw^2-2 xwx-2 xwy+xy^2\nonumber\\
 &&+4xyw-2xyx+y^2w+y^2x+yw^2\nonumber\\
 &&-2ywx-2ywy+yx^2-2yxw-2yxy).
\end{eqnarray}
In fact, in our case $w=x$ [see Eq.~\eqref{eq:square}], which results in some simplifications,
\begin{eqnarray}
 z^{(1)}&=&2x+y,\nonumber\\
 z^{(2)}&=&0,\nonumber\\
 z^{(3)}&=&\frac{1}{6}(-x^2y+xy^2+2 xyx+y^2x-yx^2-2 yxy)\nonumber\\
 &=&\frac{1}{6}([[x,y],x]+[[x,y],y]),\nonumber\\
 z^{(4)}&=&0. 
\end{eqnarray}
In particular, all even order terms vanish ($z^{(2n)}=0$).

We can now substitute $x=-iH_+ T/4$, $y=-iH_- T/2$, and $z=-iH_FT$, where
\begin{eqnarray}
 &&H_{\pm}=\Omega\sum_iP_{i-1}X_iP_{i+1}-(\Delta_0\pm\Delta_m)\sum_in_i.
\end{eqnarray}
Now we have
\begin{eqnarray}
 e^{-iH_FT} &=& e^{-iH_+\frac{T}{4}}e^{-iH_-\frac{T}{2}}e^{-iH_+\frac{T}{4}}\nonumber,\\
 -iH_FT &=& \left(-2iH_+\frac{T}{4}-iH_-\frac{T}{2} \right) \nonumber\\
 \nonumber  &+& \frac{1}{6} \left(\frac{iT^3}{32}[[H_+,H_-],H_+] 
 + \frac{iT^3}{16}[[H_+,H_-],H_-] \right). \\
\end{eqnarray}
Using the expression for the commutator,
\begin{eqnarray}
 [H_+,H_-] = -2i\Omega\Delta_m\sum_iP_{i-1}Y_iP_{i+1},
\end{eqnarray}
after some algebra, we obtain
\begin{eqnarray}\label{eq:bch_final}
\relax[[H_+,H_-], H_{\pm}] &=& -4\Omega^2\Delta_m\sum_iP_{i-1}Z_iP_{i+1} \nonumber\\
&-& 4\Omega^2\Delta_m\sum_iP_{i-1}(\sigma^+_{i}\sigma^-_{i+1}+\sigma^-_{i}\sigma^+_{i+1})P_{i+2} \nonumber\\
&+& 2\Omega\Delta_m(\Delta_0\pm\Delta_m)\sum_iP_{i-1}X_iP_{i+1},
\end{eqnarray}
which finally leads to the expression Eq.~\eqref{eq:hF} in the main text.
We see from Eq.~(\ref{eq:bch_final}) that, in addition to the PXP and detuning terms, two new terms are generated in the third order expansion: a diagonal PZP term and an off-diagonal hopping term. The ``time-crystal-like" dynamics from the N\'eel state is generated by the interplay of such terms.
The resulting Hamiltonian is sufficiently local and 
captures the subharmonic fidelity revival quite well.

At higher frequencies, the agreement between BCH expansion and full dynamics under $H(t)$ is systematically improved. At lower frequencies, one needs to go higher order in the expansion to ensure a good agreement. We note that although new non-local terms (with support over larger clusters of sites) will proliferate at higher orders, the 
terms in Eq.~(\ref{eq:bch_final}) will also occur repetitively, thus renormalizing their strength. For example, if we expand the high amplitude expression
of $H_F^{(1)}$ from Eq.~\eqref{eq:floq_ham} we get 
\begin{eqnarray}
 \frac{\Delta_0}{2\Omega}g&=&\frac{\Delta_0}{\Delta_m^2-\Delta_0^2}
 \left\{\left[1+2\frac{\sin\frac{(\Delta_m-\Delta_0)T}{4}}{\sin\frac{\Delta_0T}{2}}\right]\Delta_m-\Delta_0\right\}\nonumber\\
 &=& 1 + \frac{\Delta_mT^2(3\Delta_0-\Delta_m)}{96}\nonumber\\
 &+&\frac{\Delta_m T^4 \left(75 \Delta_0^3-7 \Delta_0^2 \Delta_m-15 \Delta_0\Delta_m^2+3\Delta_m^3\right)}{92160}\nonumber\\
 &+&\mathcal{O}(T^6).
\end{eqnarray}
Note that the first two terms are exactly same as in Eq.~\eqref{eq:hF}. Thus to get a better Floquet effective Hamiltonian, one can just re-sum the terms in Eq.~\eqref{eq:hF}. Numerically, one can tune the coefficients of the 
terms in Eq.~\eqref{eq:hF} manually to obtain a better matching between the fidelity dynamics calculated using $H_F$ and $H(t)$. Ultimately, however, the BCH expansion is expected to diverge at lower frequencies, which is at odds with the numerics that shows the existence of long-lived subharmonic response at frequencies $\omega\sim 1$. We leave the interesting question how to adapt the BCH expansion to describe this non-perturbative regime for future work.

 \section{A different way of visualising the Hilbert space trajectory}\label{a:trajectory_polarized}
 
 In the main text, we analyzed the trajectory of a driven QMBS system in the Hilbert space by evaluating the average number of excitations of the two sublattices for the $\ket{\mathbb{Z}_2}$ initial state. This method is less useful for the $\ket{0}$ state since in this case the Hilbert space trajectory in just a diagonal line $\langle n_A\rangle$ = $\langle n_B\rangle$. A more generic method for visualising the trajectory is to compute the reduced density matrix for a small subsystem, e.g.,  $\rho_2$ for a 2-site subsystem, and plot its eigenvalues. Due to the PXP constraints, there are only three possible configurations for two sites ($\bullet\circ$, $\circ\circ$ and $\circ\bullet$), thus $\rho_2$ has three eigenvalues $\lambda_i$. However, only two of those are independent since $\mathrm{tr}(\rho_2)=1$, which allows us to plot the trajectory in 2D. 

 \begin{figure}[tb]
 \includegraphics[width=0.45\textwidth]{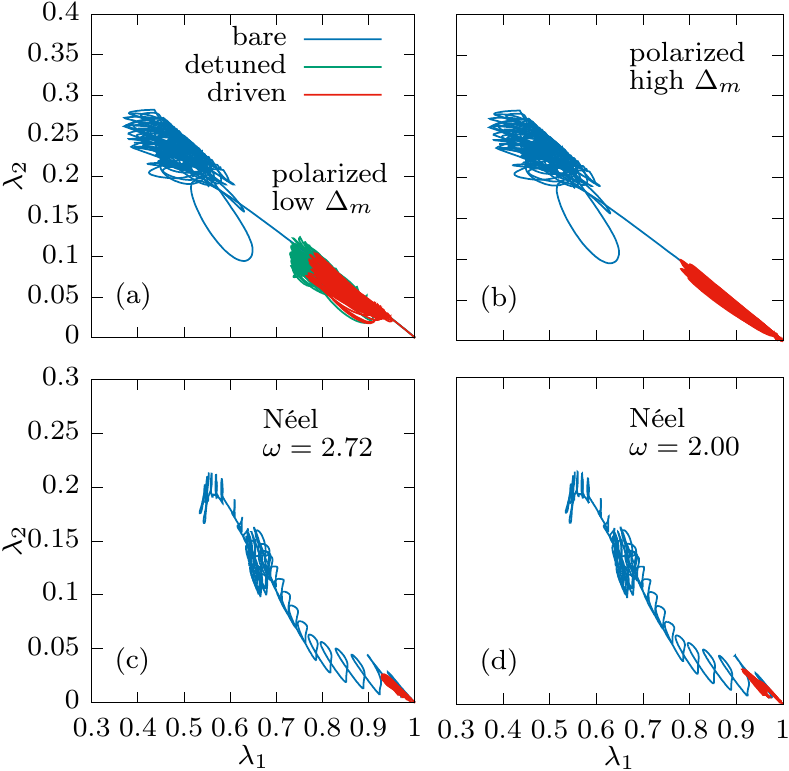}
 \caption{Visualising the Hilbert space trajectory using the 2-site reduced density matrix eigenvalues. System size $N=16$, without driving [$\Delta(t)=0$, blue], with static detuning [$\Delta(t)=\Delta_0$, green] and with optimal cosine drive (red). (a) Polarized state, $\Delta_0=1.68$, $\Delta_m=-0.50$, $\omega=3.71$. (b) Polarized state, $\Delta_0=0.64$, $\Delta_m=7.55$, $\omega=2.90$. (c)  N\'eel state, $\Delta_0=1.15$, $\Delta_m=2.67$, $\omega=2.72$. (d)  N\'eel state, $\Delta_0=-1.17$, $\Delta_m=4.81$, $\omega=2.00$. 
 }\label{fig:trajectory_2site}
 \end{figure}

 The trajectories of the polarized state in two different optimal driving regimes can be observed in Figs.~\ref{fig:trajectory_2site}(a) and \ref{fig:trajectory_2site}(b). Here we use the largest and the smallest eigenvalue of $\rho_2$ and compare the cases without driving (bare PXP model), with static detuning only, and with periodic driving. The polarized state is located at the $(1,0)$ point (bottom right corner), since its reduced density matrix corresponds to a pure state ($\lambda_1=1$, $\lambda_2=\lambda_3=0$).
 Similar to the N\'eel state, the driven trajectory approximately repeats the first part of the static trajectory. There are no revivals from the polarized state in the bare PXP model, so the trajectory at later times ends up in the region where all three eigenvalues are of approximately equal magnitudes, implying thermalization. In the low-amplitude regime, the static detuning alone is enough to produce revivals, as evidenced by the trajectory, which periodically returns to the vicinity of the initial state, while periodic driving further stabilizes the detuned trajectory, see Fig.~\ref{fig:trajectory_2site}(a). In the high-amplitude regime, the revivals are created purely by driving. In that case the driven trajectory is even more narrow and returns closer to the initial state, see Fig.~\ref{fig:trajectory_2site}(b).

For comparison, we use the same method to plot the trajectories for the N\'eel state in Figs.~\ref{fig:trajectory_2site}(c) and \ref{fig:trajectory_2site}(d). In these two figures we use two different sets of driving parameters, one of which corresponds to the ideal trajectory with no loops in Fig.~\ref{fig:trajectory_neel}(c) and the other to a ``one-loop'' trajectory from Fig.~\ref{fig:trajectory_neel}(b). However, it is now much harder to see the difference between these two cases. In both cases the largest eigenvalue stays close to 1.

\end{document}